\documentclass[11pt,titlepage]{article}
\usepackage[utf8]{inputenc} 
\usepackage{graphicx}
\graphicspath{{figures/}}
\usepackage{longtable}
\usepackage[round]{natbib}
\usepackage{pseudocode}
\usepackage{amsmath}
\usepackage{amsfonts}
\usepackage{amssymb}
\usepackage{ntheorem}
\theoremseparator{:}

\usepackage{multirow, array}
\usepackage{bigstrut}
\usepackage{lscape}
\usepackage{float}
\usepackage{todonotes}
\usepackage{booktabs}
\usepackage[margin=1in]{geometry}

\long\def\symbolfootnote[#1]#2{\begingroup%
\def\thefootnote{\fnsymbol{footnote}}\footnote[#1]{#2}\endgroup}
\usepackage{caption}
\usepackage{subcaption}

\usepackage{changepage}
\usepackage{color,hyperref}
\definecolor{darkblue}{rgb}{0.0,0.0,0.3}
\hypersetup{colorlinks,breaklinks,
            linkcolor=darkblue,urlcolor=darkblue,
            anchorcolor=darkblue,citecolor=darkblue}

\begin{document}

\begin{titlepage}

\noindent {\bf \Large $S$-maup: Statistic test to measure the sensitivity to the \\ \\Modifiable Areal Unit Problem}\\[1.5cm]

\noindent {\bf Juan~C.~Duque }\\
Department of Mathematical Sciences (RiSE-group), Universidad EAFIT, Medellín, Colombia. \\
 E-mail:\texttt{jduquec1@eafit.edu.co}\\

\noindent {\bf Henry~Laniado}\\
Department of Mathematical Sciences (RiSE-group), Universidad EAFIT, Medellín, Colombia. \\
 E-mail:\texttt{hlaniado@eafit.edu.co}\\

\noindent {\bf Adriano~Polo }\\
Department of Economics, Universidad EAFIT, Medellín, Colombia. \\
 E-mail:\texttt{apololo@eafit.edu.co}\\

\noindent {\bf Abstract:}

This work presents a nonparametric statistical test, $S$-maup, to measure the
sensitivity of a spatially intensive variable to the effects of the
Modifiable Areal Unit Problem (MAUP). $S$-maup is the first statistic of
its type and focuses on determining how much the distribution of the
variable, at its highest level of spatial disaggregation, will change when
it is spatially aggregated.  Through a computational experiment, we obtain
the basis for the design of the statistical test under the null hypothesis
of non-sensitivity to MAUP.  We performed a simulation study for
approaching the empirical distribution of the statistical test, obtaining
its critical values, and computing its power and size. The results
indicate that the power of the statistic is good if the sample (number of
areas) grows, and in general, the size decreases with increasing sample
number.  Finally, an empirical application is made using the Mincer
equation in South Africa.

\end{titlepage}

\section{Introduction}
\label{sec:intro}

Although spatial data are increasingly disaggregated, many socioeconomic
studies require some level of aggregation (e.g., neighborhoods,
municipalities, states, districts, countries). Spatial aggregation is useful
for calculating rates and indexes, minimizing the influence of outliers, or
preserving confidentiality \cite{wise:97,wise:01}. Spatial aggregation is also useful for
creating meaningful units for analysis \cite{yule:1950,Duque2006}, reducing
computational complexity \cite{miller:99}, controlling for spurious spatial
autocorrelation \cite{bian:99, Duque2012}, and comparing results at different
scales \cite{holtarea:1996, tagashira:02}.

However, spatial aggregation triggers a problem known as the Modifiable Areal
Unit Problem (MAUP). The MAUP, introduced in the literature by
\cite{openshaw:1978} and \cite{openshaw:1979}, refers to the sensitivity of
statistical results to changes in the spatial units of analysis. The MAUP has
two dimensions: the scale effect and the zoning effect. The scale effect
refers to changes in the size of the spatial units, which implies a change in
the number of spatial units, e.g., doing the analysis at the state or county
level. The zoning effect refers to changes in the shape of the spatial units
preserving the number of units, e.g., aggregating USA counties into 50
states is merely one of the many ways in which one can aggregate counties into
50 spatial units.

Although the literature on MAUP is extensive, to the best of our knowledge,
there is no statistical tool that allows a practitioner to easily determine
the level of sensitivity of a spatially intensive variable to the MAUP. Hence,
in this paper, we present $S$-maup, a nonparametric statistical test to measure
the sensitivity of a spatially intensive variable to the MAUP.  Instead of
looking at a specific measure of central tendency or dispersion or at the
coefficient associated with the variable in a specific regression, $S$-maup
focuses on determining how much the distribution of the variable, at its
highest level of spatial disaggregation, will change when it is aggregated
into a given number of regions. For its calculation $S$-maup requires the
number of areas, the $\rho$ parameter, that measures the degree of spatial
correlation of the variable, and the number of regions in which the areas will
be aggregated. Under the null hypothesis of non-sensitivity to MAUP, $S$-maup
would be useful for determining the maximum level of aggregation that we can
apply to a given variable before it loses its distributional characteristics.
$S$-maup could also be used to determine whether the results obtained at a
given scale (e.g., counties) hold for another scale (e.g., states).

The rest of this article is structured as follows. We begin with a literature
review concerning the primary research surrounding the MAUP. We then explore the effects of
the MAUP through a computational experiment. Next, we propose a test
statistic, $S$-maup, and its empirical distribution under the null hypothesis of
non-sensitivity to MAUP.  Next, we establish the power and size of the
statistic under various levels of spatial autocorrelation and number of areas.
We then present a simple example of the use of the $S$-maup statistic. Last,
we conclude and suggest avenues for further investigation.

\section{Literature Review}
\label{sec:review}

The effects of aggregating spatial data have been a subject of study since the
early 1930s and have been referred to by different names, such as
aggregation effects \cite{gehlke:1934}, scale problem \cite{yule:1950},
ecological fallacy \cite{robinson:1950}, and Modifiable Areal Unit Problem,
MAUP, \cite{openshaw:1979}. If one delves into the details, it can be
argued that these previous concepts are different. However, these concepts possess as a
common factor a concern regarding the undesired effects that result from
working with aggregate data. Hereinafter, we will refer to this problem as
MAUP.

The literature on MAUP can be divided into three blocks: first, definition of
the problem \cite{openshaw:1977, openshaw:1979, arbia:1989}; second,
measurement of its effects on statistics such as the mean \cite{amrhein:1995,
steelrules:1996}, median and standard deviation \cite{bian:99}, variance and
covariance \cite{amrheinRey:1996, reynolds:1998}, and correlation coefficient
\cite{gehlke:1934, yule:1950, openshaw:1979, clark:1976}; and last, potential
ways to minimize the aggregation effects \cite{coulson:1978,
fotheringham:1989, arbia:1989, fotheringham:2000, carrington:06}.

It is well known that the impact of the MAUP on the mean can be considered
negligible \cite{arbia:1989, amrhein:1995, amrheinRey:1996,
steelrules:1996}. However, the MAUP has a large impact on the variance, which
decreases when the variable exhibit high values of spatial autocorrelation
\cite{reynolds:1998}. With respect to the statistical association, such as the
covariance and correlation coefficient, \cite{clark:1976},
\cite{openshaw:1979} and \cite{arbia:1989} found that the sensitivity to
MAUP increases as the level of spatial aggregation increases (scale effect),
i.e., the correlation between variables $X$ and $Y$ will exhibit a wider
variation if, for example, USA counties are aggregated into 50 spatial units
than if they were aggregated into 1,000 spatial units.

The MAUP effects have also been studied in OLS regressions \cite{clark:1976,
openshaw:1978, green:1996, tagashira:02}, logit models
\cite{fotheringham:1991}, Poisson regression \cite{flowerdewP:1989}, spatial
interaction models \cite{arbia:2013}, spatial econometrics models
\cite{arbia:2011}, forecasts in regional economy \cite{miller:1998}, and
spatial autocorrelation statistics, such as the Moran's coefficient, Geary's
Ratio, and G-Statistic \cite{fotheringham:1991, qi:1996, jelinski:1996}. Other
authors have studied the MAUP effects in more sophisticated methods, such as
the factorial analysis \cite{hunt:1996}, spatial interpolation
\cite{cressie:1996}, image classification \cite{arbia:1996}, location and
allocation models \cite{goodchild:1979}, and discrete selection models
\cite{guo:2004}.

Although there is no solution to the MAUP because it is inherent to the use
of spatial data, some authors have proposed different alternatives to minimize
its effects: the formulation of scale-robust statistics \cite{king:1997}, the
design of optimal aggregations that minimize the loss of information
\cite{moellering:1972, openshaw:1977, nakaya:2000, tagashira:02, Duque2006},
the use of a set of auxiliary or grouping variables together with variables at the
individual level \cite{holt:1996, wrigleyb:1996}, and the measurement of
rates of change through the concept of a fractal dimension
\cite{fotheringham:1989}.

Most studies above required extensive computational experiments. Table
\ref{tab:resumen} summarizes the main characteristics of those experiments,
including the covered dimensions (scale or zoning), studied statistics (mean,
variance, correlation, regression coefficients, etc.), type of data (real or
simulated), studied variables (income, rates, random, etc.), and size of the
experiment in terms of the number of areas and regions (herein, we will
refer to area as the smallest spatial unit of observation and region as the
spatial units that result from aggregating the areas into contiguous spatial
units). From this table, we can highlight the dominance of the use of real data
over simulated data and the evident increase in the size of the experiment as
the computational capacity increases over the years. As expected, the two
driver parameters in these experiments are the number of areas and the number
of regions. Although it has been considered in a few experiment
\cite{openshaw:1979, reynolds:1998, bian:99}, the level of spatial
autocorrelation of the variables/attributes being aggregated plays an
important role in the level of sensitivity of the variable to the MAUP.
Finally, the mean is significantly highlighted by being the more common grouping
operator, i.e., if areas $i$ and $j$, with attribute values $X_i$ and $X_j$,
are merged into a region, the attribute value for the resulting region is
calculated as the mean of $X_i$ and $X_j$, which indicates that all of the
experiments use spatially intensive variables.

Based on the available literature, a practitioner can anticipate high(low)
variation of its results when the aggregation level is high(low) and the level
of spatial autocorrelation of its variable is low(high). However, there is no tool
in the literature that allows the assignment of a specific number and statistical
significance to that variation. The closest the research can get to that
number would require a computational experiment involving the calculation
of the results for a large number of random aggregations of the areas into a
predefined number of regions. This paper constitutes the very first attempt to
formulate a nonparametric statistical test to easily measure the sensitivity
of a spatially intensive variable to the MAUP.

\begin{table}
\begin{adjustwidth}{-0.60in}{0in} 
    \centering
    \caption{{\bf Computational experiments on MAUP.}}
    \scalebox{0.6}{
    \begin{tabular}{p{2.5 cm} p{2.2 cm} p{2.7 cm} p{3.5 cm} p{5.7 cm} p{8.2 cm}}
    \toprule
    \textbf{Author (Year)} & \textbf{Dimension / Effect on...} &
    \textbf{Grouping operator} & \textbf{Data} & \textbf{Variable}
    & \textbf{Size} \\
    \midrule
    \cite{gehlke:1934} & Scale / $\displaystyle \ r_{xy}$    &
    Sum &Census Tracts in Cleveland & Male juvenile delinquency
    and monthly income. Agricultural products and the number of
    farmers & 1) 252 areas into 200, 175, 150, 125,
    100, 50, and 25 regions 2) 1,000 areas into 63,
    40, 31, and 8 regions \\
    \midrule
    \cite{robinson:1950} & Scale / $\displaystyle \ r_{xy}$ &
    Proportions & Nine geographic divisions of the USA in 1930 &
    Race and illiteracy & 97,272 individuals into 9 regions \\
    \midrule
    \cite{yule:1950} & Scale / $\displaystyle \ r_{xy}$ & Mean &
    Agricultural counties in England & Production of wheat and
    potatoes per acre & 48 areas into 24, 12, 6, and 3 regions \\
    \midrule
    \cite{clark:1976} & Scale / $\displaystyle \ r_{xy}$ & Mean &
    Metropolitan area of Los Angeles & Household income and
    education level of the head of household & 1,556 census tracts
    into 134 Welfare Planning Council Study areas
    and 35 Regional Planning Commission Statistical Areas \\
    \midrule
    \cite{openshaw:1979} & Scale - Zoning / $\displaystyle \
    r_{xy}$ & - & Counties in Iowa and simulated data with
    $\displaystyle \rho_{+}, \rho_{0}$, and $\displaystyle \rho_{-}$
    & \% of Republican votes and \% population over 60 years. & 99
    areas into 6, 12, 18, 24, 30, 36, 42, 48, 54, 60, 66, and 72
    regions\\
    \midrule
    \cite{arbia:1989} & Scale - Zoning / $\displaystyle \mu,
    \sigma ^{2}$, $\displaystyle \sigma, \rho$ & Mean & Quadrat
    in Hukuno Town, Japan and weights of wheat
    plots of grain&  Quadrat counts of houses and weights of wheat
    plots of grain  & 1) Regular lattice of 32x32 into 16x16, 8x8,
    4x4, and 2x2 regions 2) Regular lattice of 25x20 cells into
    8x8, 4x4, and 2x2 regions \\
    \midrule
    \cite{fotheringham:1991} & Scale - Zoning / $\displaystyle
    \beta's$ & Mean and Proportion & Metropolitan area of Buffalo &
    Household income, \% of population per area, \% of population
    over 65 years & 871 areas into 800, 400, 200, 100, 50, and 25
    regions \\
    \midrule
    \cite{amrhein:1995}, \cite{steelrules:1996} & Zoning /
    $\displaystyle \mu, \sigma ^{2}$, $\displaystyle \ r_{xy}$ y
    $\displaystyle \beta$ & Mean and weighted average & Regular
    lattices & Simulated data with Uniform, Normal and Poisson distribution & 10,000
    areas into 10x10, 7x7, and 3x3 regions \\
    \midrule
    \cite{holtarea:1996} & Zoning / $\displaystyle \sigma ^{2}$ &
    Mean & City of Adelaide, Australia & 82 socioeconomic
    variables & 917,000 people into 1,584 districts \\
    \midrule
    \cite{amrheinRey:1996} & Zoning / $\displaystyle \sigma ^{2}$
    & Mean & Lancashire, UK & 8 census variables & 304 areas into
    137, 122, 106, 91, 76, 61, 46, and 30 regions \\
    \midrule
    \cite{green:1996} & Scale - Zoning / $\displaystyle \
    r_{xy}$, $\displaystyle \beta$ & Mean & UK & Census
    variables and simulated variables & Regular lattice of 120x120
    into 1x1, 2x2, 3x3, 4x4, and 5x5 regions\\
    \midrule
    \cite{qi:1996} & Scale / $I-Moran$ and $G-Statistic$ &
    Mean & Malasia & Biomass areas and elevation data & Regular
    Lattice of 220x188 into 2x2, 3x3, 4x4, ..., and 20x20 regions \\
    \midrule
    \cite{jelinski:1996} & Scale / $I-Moran$ and $G-Statistic$ &
    Mean & Manitoba, Canada & Normalized Difference Vegetation
    Index (NDVI) & Regular lattice of 300x300 into 3x3, 5x5, 7x7,
    9x9, 11x11, 13x13, and 15x15 areas \\
    \midrule
    \cite{reynolds:1998} & Zoning / $\displaystyle \sigma ^{2}$,
    $\displaystyle \ r_{xy}$, and $\displaystyle \beta$ & Mean &
    Regular lattices & Simulated variables with different levels
    of spatial autocorrelation and variance & 400 areas into 180,
    160, 140, 120, 100, 80, 60, and 40 regions \\
    \midrule
    \cite{bian:99} & Zoning / $\displaystyle \sigma$ & Mean and
    Median & Regular lattices & Simulated data with different
    levels of spatial autocorrelation & Regular lattice of 512x512
    into 3x3, 9x9, 11x11, 21x21, 31x31, 41x41, 51x51, 61x61,
    71x71, and 81x81 pixel window sizes \\
    \midrule
    \cite{arbia:2011, arbia:2013} & Scale / $\displaystyle \beta$
    & Mean & Regular lattice & Simulated data & Regular lattice of
    64x64 into 32x32, 16x16, 8x8, and 4x4. \\
    \bottomrule
    \end{tabular}
    }
    \begin{flushleft}\scriptsize{ $\displaystyle \ r_{xy}$:
    Correlation, $\displaystyle \mu$: Media, $\displaystyle \sigma ^{2}$:
    Variance, $\displaystyle \sigma$: Covariance, $\displaystyle \rho$:
    Spatial autocorrelation, $\displaystyle \beta$: Regression
        coefficients.}
    \end{flushleft}
	\label{tab:resumen}
\end{adjustwidth}
\end{table}

\section{MAUP Effects}
\label{sec:efectos}

In this section, we design a computational experiment to identify the key
elements that should be included in the construction of the statistical test.
Following previous experiments in the literature on the MAUP effects (e.g.,
\cite{amrheinRey:1996} and \cite{arbia:2011}), we consider the two main
parameters involved in the exploration of scale and zoning effects: number of
areas ($N$) and number of regions ($K$). As in \cite{openshaw:1979} and
\cite{reynolds:1998}, we also take into account different levels of spatial
autocorrelation, $\rho$.

\begin{figure}
    \centering
	\includegraphics[width=.6\textwidth]{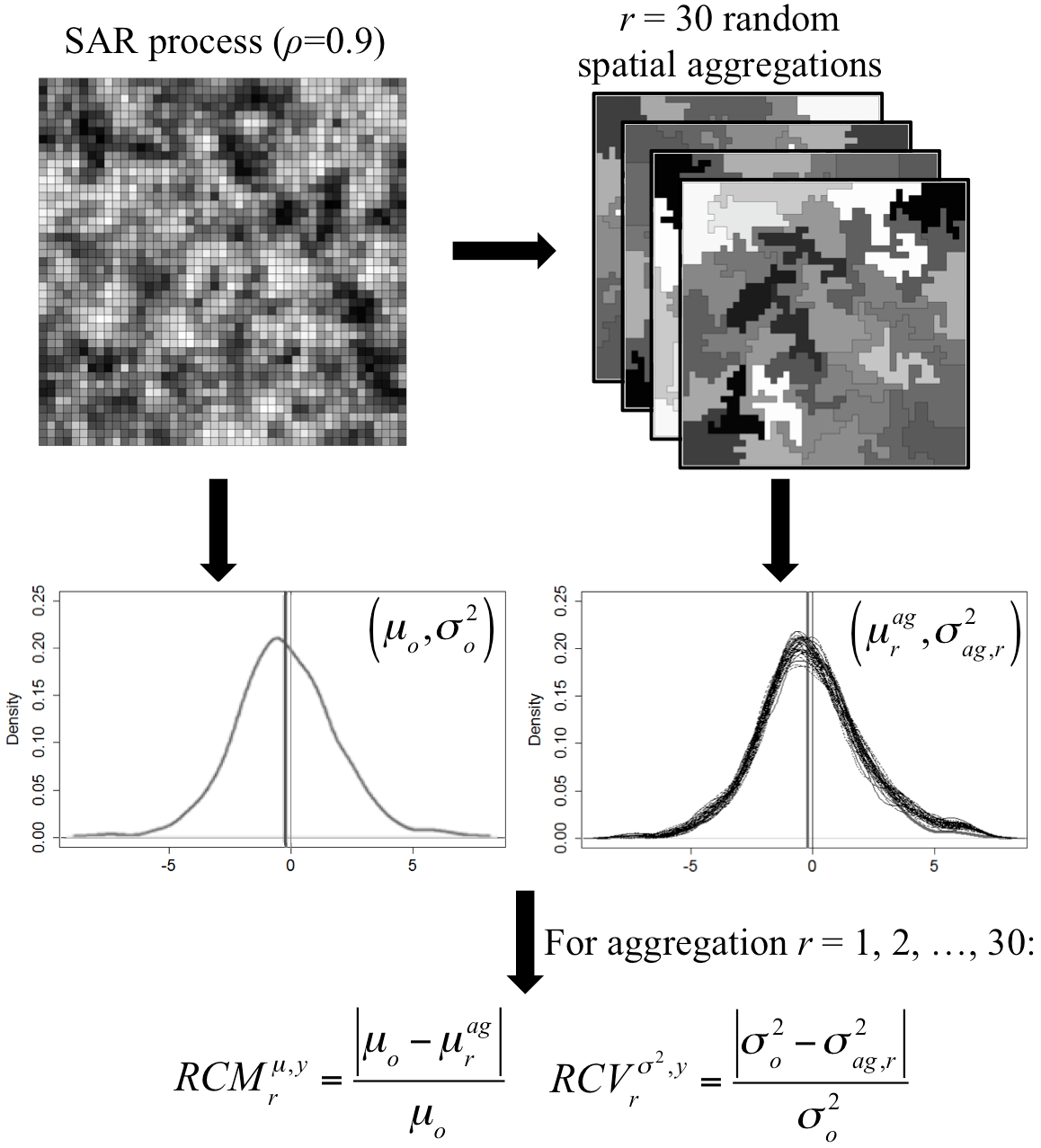}
	\caption{Instance of the experiment.}
    \label{instance}
\end{figure}

Fig \ref{instance} summarizes the steps followed to generate an instance of
the experiment: (1) $y^{\rho=0.9}$ is a random variable generated by a Spatial
Autoregressive (SAR) process with autoregressive parameter $\rho = 0.9$ and
rook contiguity matrix. (2) The areas are randomly aggregated into $K$
spatially contiguous regions using a seed-based region growing algorithm
proposed by \cite{vickrey:1961}. The attribute value for each region is
calculated as the mean value of the attribute values of the areas assigned to
the region. This random aggregation is repeated $r = 30$ times, so that we
generate $30$ different ways to aggregate $N$ areas into $K$ regions. (3) We
calculate the mean and variance of the original, disaggregated, variable as
$\mu_{o}$ and $\sigma_{o}^{2}$. (4) We calculate the mean and variance of each
one of the aggregated variables as $\mu_{ag}$ and $\sigma_{ag}^{2}$. (5) We
calculate the relative change in the mean ($RCM$), Eq \eqref{eq:1}, and
the relative change in the variance ($RCV$), Eq \eqref{eq:2}, between the
original variables and each of the 30 aggregated variables. (6) We summarize
the effect of aggregating $N$ areas into $K$ regions as the mean $RCM$,
$\overline{RCM}$, and mean $RCV$, $\overline{RCV}$, using Eq \eqref{eq:3}
and \eqref{eq:4}. For each value of $\rho$ considered in the experiment, we
repeat steps (1) to (6) 50 times.

\begin{equation}
    RCM_{r}^{\mu,y} = \frac{\left |\mu_{o} - \mu_{r}^{ag} \right |}{\mu_{o}}
    \label{eq:1}
\end{equation}

\begin{equation}
    RCV_{r}^{\sigma^2,y} = \frac{\left |\sigma_{o}^{2} - \sigma_{ag,r}^{2} \right |}{\sigma_{o}^{2}}
    \label{eq:2}
\end{equation}

\begin{equation}
    \overline{RCM} =\frac{\sum_{r=1}^{30} RCM_{r}^{\mu,y}}{30}
    \label{eq:3}
\end{equation}

\begin{equation}
    \overline{RCV} =\frac{\sum_{r=1}^{30} RCV_{r}^{\sigma^2,y}}{30}
    \label{eq:4}
\end{equation}

\begin{figure}
    \centering
	\includegraphics[width=0.99\textwidth]{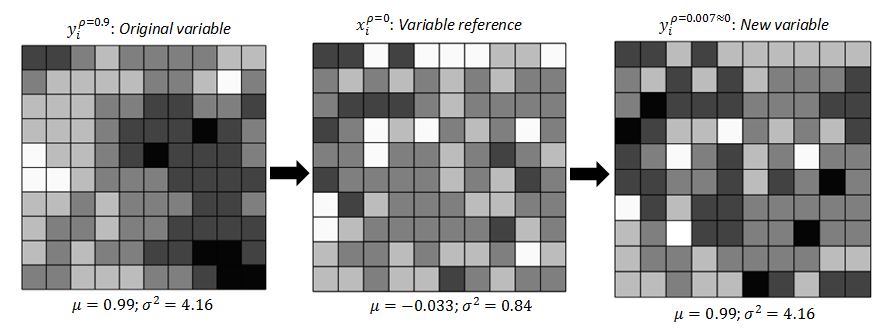}
	\caption{Example of spatial autocorrelation generation.}
    \label{f_rhos}
\end{figure}

As we will show in the parametrization of the experiment, we generate
instances of $y^{\rho}$ for different levels of spatial autocorrelation
($-0.9<\rho<0.9$). If, for example, we generate two spatial processes
$y^{\rho=0.9}$ and $y^{\rho=-0.5}$, the observed aggregation effects will have
two sources, one that comes from the change in the value of $\rho$, and one
that comes from the differences in the values generated by the random data
generation process. To isolate the effect that comes from the changes
in $\rho$, we generate the instances of $y^{\rho=0.0}$ by performing spatial
permutations of the values obtained from $y^{\rho=0.9}$. As an example, Fig
\ref{f_rhos} summarizes the process that we implemented to generate
$y^{\rho=0.0}$ from $y^{\rho=0.9}$: (1) Generate an SAR process $y^{\rho=0.9}$.
(2) Generate a reference SAR process $x^{\rho=0.0}$. (3) Generate
$y^{\rho=0.0}$ by spatially redistributing the values of $y^{\rho=0.9}$
following the spatial pattern of $x^{\rho=0.0}$, i.e., the highest value of
$y^{\rho=0.9}$ goes to the area with the highest value of $x^{\rho=0.0}$; the
second highest value of $y^{\rho=0.9}$ goes to the area with the second
highest value of $x^{\rho=0.0}$; and so forth. (4) Estimate the true $\rho$
value of $y^{\rho=0.0}$, if $(0.0-0.5)<\rho<(0.0+0.5)$ then keep
$y^{\rho=0.0}$; otherwise, repeat the process. Note that $y^{\rho=0.9}$ and
$y^{\rho=0.0}$ have the same values and therefore the same mean and variance,
But due to the differences in the spatial distribution of the values,
they have different $\rho$ values.

Having clarified the process that we follow at each instance and our strategy for
generating the $y^{\rho}$ values, we present the parameters used in the
computational experiment:

$\begin{array}{ll}
	N =& \mbox{Number of areas. } N = \left\{25,100,225,400,625,900 \right\};\\
	
    y_{i}^{\rho} =& \mbox{SAR process with } i = \left\{ 1,\dots,50
    \right\}\mbox{, and } \rho = \left\{\pm 0.9, \pm 0.7, \pm 0.5, \pm 0.3, 0
    \right\};\\

    k =& \left\{ \begin{array}{l} \mbox{for $N$=25, } k= \left\{3, 5, 10, 13,
    15, 18, 20, 22, 24 \right\} \\ \mbox{for $N$=100, } k=\left\{2, 4, 7, 12,
    25, 40, 53, 67, 80, 90, 99 \right\} \\ \mbox{for $N$=225, } k=\left\{3, 5,
    10, 15, 30, 60, 90, 120, 150, 180, 200, 220\right\} \\ \mbox{for $N$=400,
    } k=\left\{4, 9, 18, 26, 50, 110, 160, 213, 267, 320, 360, 396 \right\} \\
    \mbox{for $N$=625, } k=\left\{4, 6, 14, 27, 43, 80, 170, 250, 333, 417,
    500, 563, 618 \right\} \\ \mbox{for $N$=900, } k=\left\{4, 9, 20, 40, 60,
    120, 240, 360, 480, 600, 720, 810, 890 \right\} \end{array} \right. \\

	r = 30& \mbox{Number of random spatial aggregations.}  \\
\end{array}$\\

We implemented the experiment in Python 2.7.10. For the spatial aggregations,
we use the Python library ClusterPy 0.9.9 \cite{ClusterPy}. We ran the
experiment in the supercomputer \textit{APOLO}, at the Center of Scientific
Computation (Universidad EAFIT), equipped with a Dell Power Egde 1950 III of 8
cores, 2.33 GHz Intel Xeon that executes Linux Rocks 6.1 to 64 bits.

\begin{figure}
    \centering
	\includegraphics[width=0.99\textwidth]{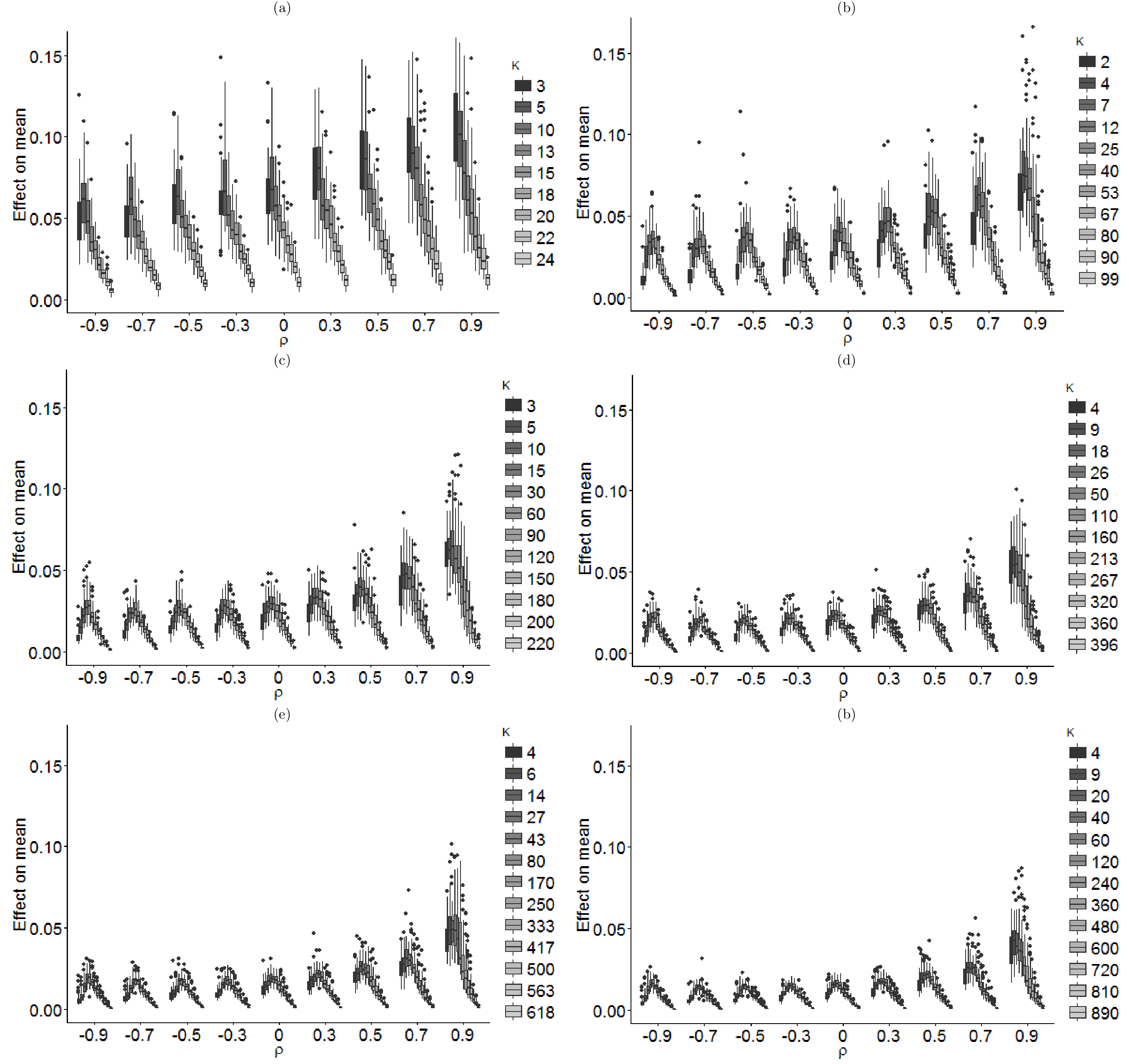}
	\caption{Relative change in mean - Average effect. (a) $N=25$; (b) $N=100$; (c) $N=225$; (d) $N=400$; (e) $N=625$; (f) $N=900$.}
    \label{fig:efm}
\end{figure}

Each box plot in Fig \ref{fig:efm} summarizes the 50 values of
$\overline{RCM}$ calculated for each value of $\rho$ and $k$. The maximum
bounds value of the vertical axis in the figure show low relative changes in the
mean. To make sure that the mean effect can be discarded, we calculate the
two-sample $t$-test to compare the mean of each original variable, $\mu_{o}$,
with the mean of each aggregated variable, $\mu_{ag}$. We report in Table
\ref{tab:efm} the proportion of instances for which the two-sample $t$-test
was rejected. From this result, we can conclude that there is not a MAUP effect
on the mean, which is consistent with those results found by
\cite{arbia:1989, amrhein:1995} and \cite{amrheinRey:1996}.

\begin{table}
    \centering
    \caption{{\bf Effect on mean - Proportion of significant instances.}}
    \begin{tabular}{ccccccc}
    \toprule
    & \multicolumn{6}{c}{Number of areas}   \\
    &    $N=25$  & $N=100$  & $N=225$   & $N=400$ & $N=625$& $N=900$\\
    \midrule
    Proportion* & 0  & 0.00063 & 0.00014 & 0.00041 & 0.00063 & 0.0012\\
    \bottomrule
    \end{tabular}
    \begin{flushleft}\footnotesize{ Proportion of instances for which the
    two-sample $t$-test was rejected with $\alpha=0.05$. It includes
    instances with $k\geq 10$.}
    \end{flushleft}
    \label{tab:efm}
\end{table}

\begin{figure}
    \centering
	\includegraphics[width=0.99\textwidth]{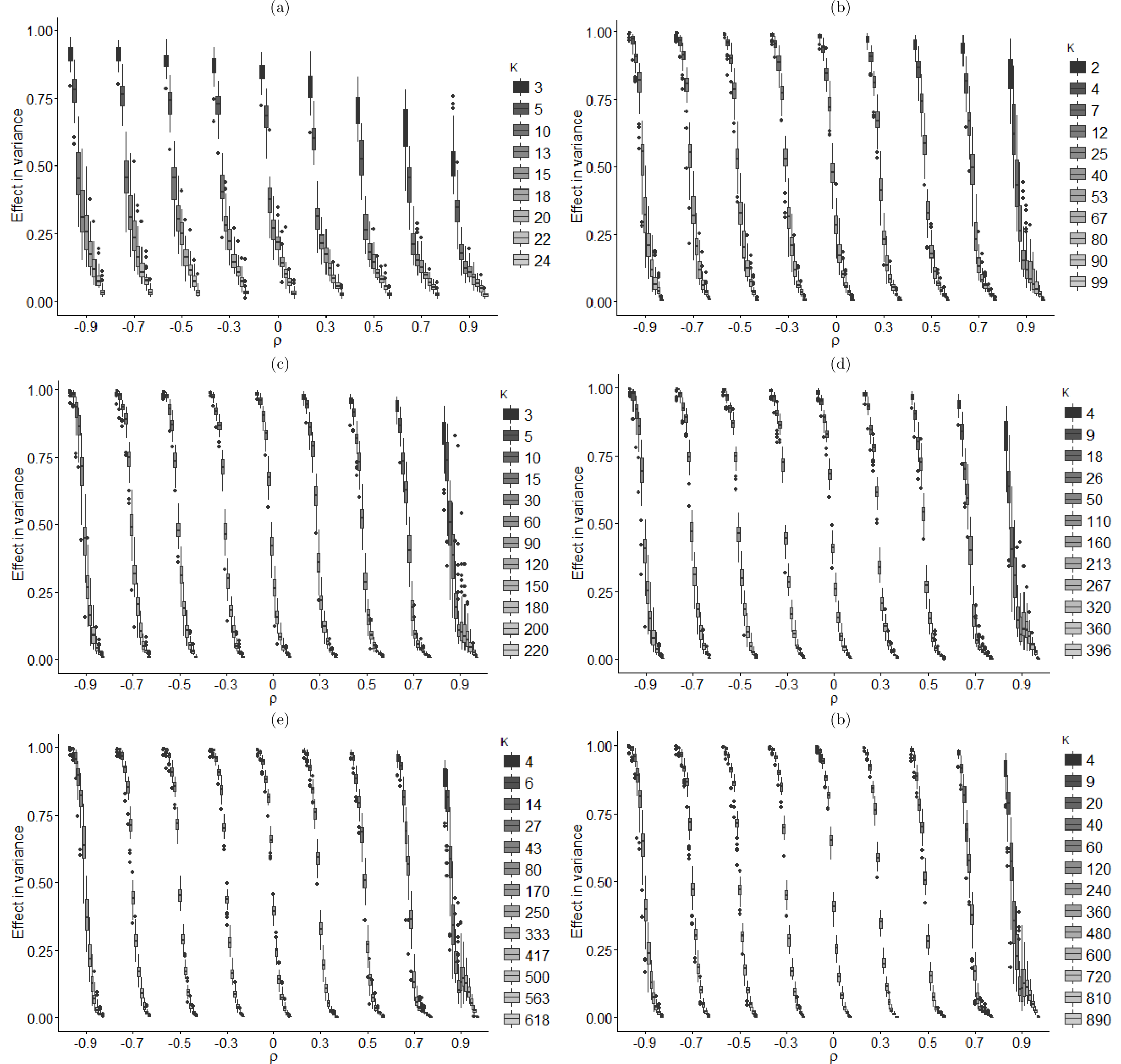}
	\caption{Relative change in variance - Average effect. (a) $N=25$; (b) $N=100$; (c) $N=225$; (d) $N=400$; (e) $N=625$; (f) $N=900$.}
    \label{fig:efv}
\end{figure}

Each box plot in Fig \ref{fig:efv} summarizes the $i = 50$ values of
$\overline{RCV}$ calculated for each value of $\rho$ and $k$. Unlike the case
seen with the mean, the effect of variance is considerably greater. The box
plots show that the effect of MAUP on variance decreases for two reasons: an
increase in the level of spatial autocorrelation, $\rho$; and (2) an
increase in the number of regions, $k$. These effects on variance are
consistent with those found by \cite{reynolds:1998}.

\begin{figure}
    \centering
	\includegraphics[width=0.99\textwidth]{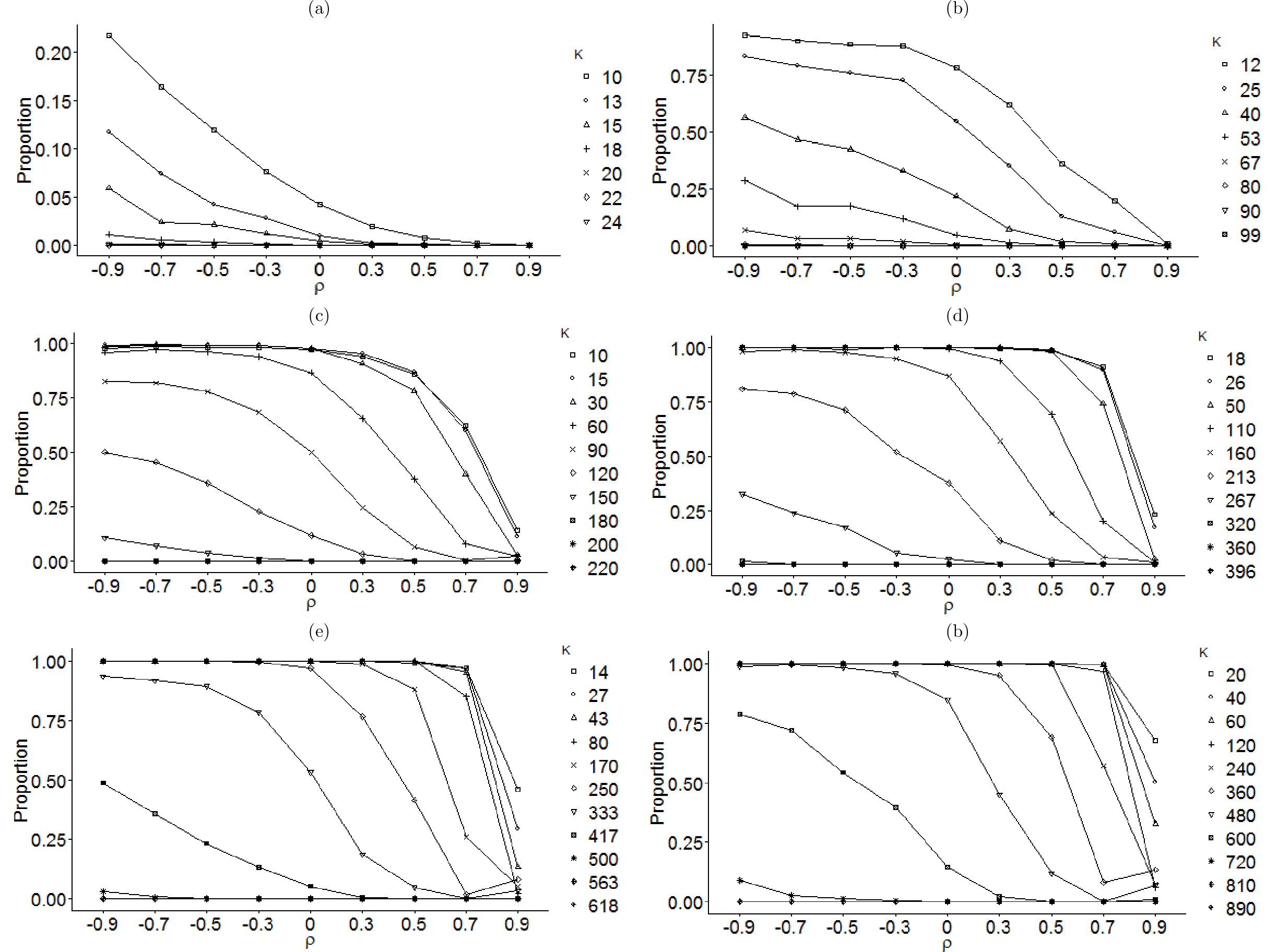}
	\caption{Proportion of instances for which the Levene test rejects the null
    hypothesis of equality of variance, with a level of significance $\alpha=0.05$. (a) $N=25$; (b) $N=100$; (c) $N=225$; (d) $N=400$; (e) $N=625$; (f) $N=900$.}
    \label{fig:levene}
\end{figure}

To verify the MAUP effect on variance, we use the Levene test for equality
between the variance of the original variable, $\sigma_{o}^{2}$, with the
variance of each aggregated variable, $\sigma_{ag}^{2}$. Fig
\ref{fig:levene} shows the percentage of instances for which the Levene test
rejects the null hypothesis $H_0:\;\sigma_{o}^{2} = \sigma_{ag}^{2}$, with
$\alpha=0.05$. These results confirm that the MAUP effect decreases as either
$k$ or $\rho$ increases.

\begin{figure}
    \centering
	\includegraphics[width=0.99\textwidth]{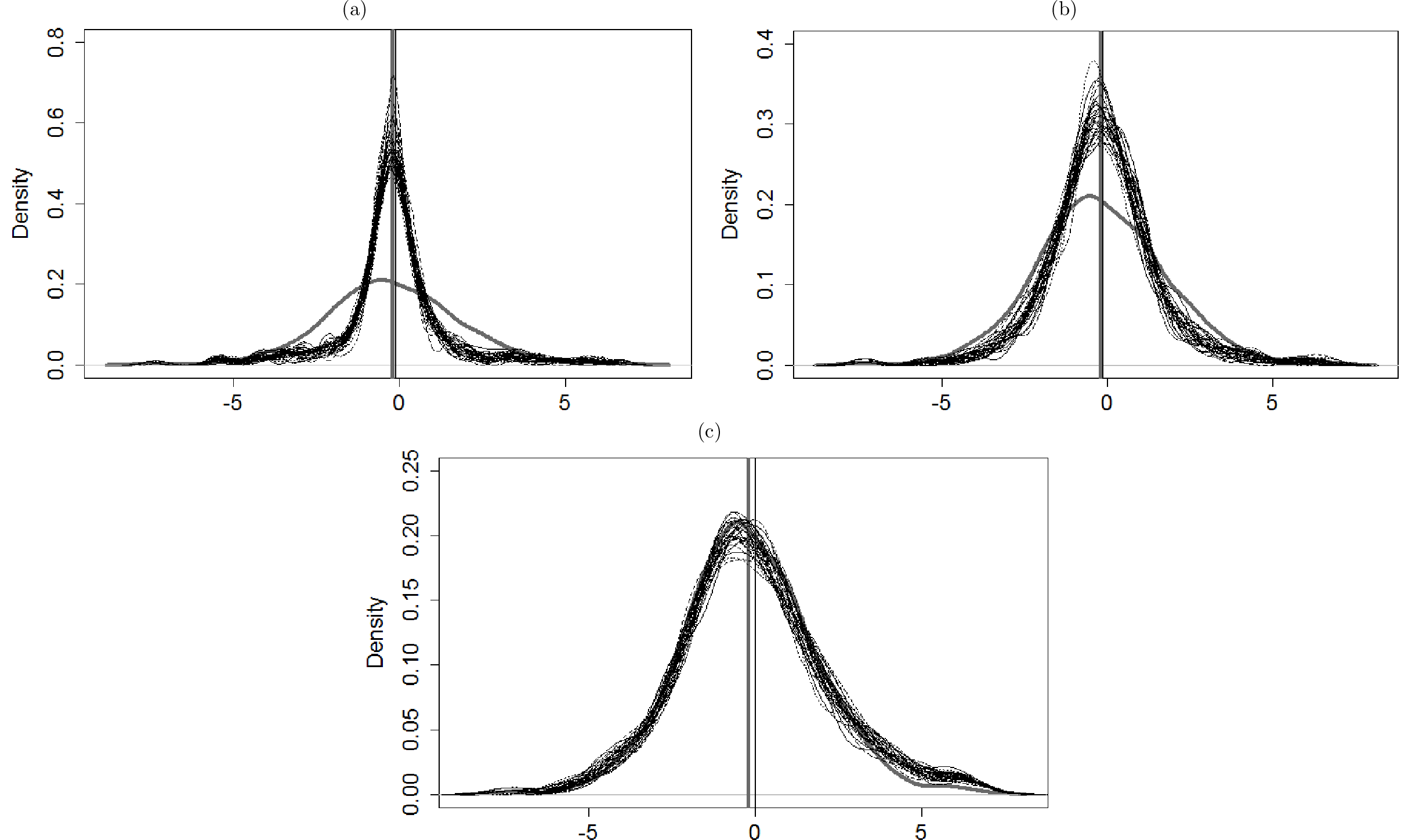}
	\caption{MAUP effects at three levels of spatial autocorrelation, (a)
    $\rho=-0.9$, (b) $\rho=0$, and (c) $\rho=0.9$. Solid line: original
    variable with $N=900$; dashed lines: 30 aggregations with $k=240$. The
    vertical lines indicate $\mu_{o}$ and $\mu_{ag}$.}
    \label{grap:ker}
\end{figure}

Finally, in Fig \ref{grap:ker} we present, for illustrative purposes, three
instances with $\rho=-0.9$, $\rho=0.0$ and $\rho=0.9$ that aggregate $N=900$
areas into $k=240$ regions. These examples show how the MAUP fades as $\rho$
increases.

\section{ $S$-maup statistical test}
\label{sec:test}

Findings such as the effect of MAUP on variance and how MAUP fades as $\rho$
and $k$ increase are useful to find the functional form of our statistical
test, $S$-maup, for measuring the level of sensitivity of a spatially
distributed variable to the MAUP. We designed the test such that $S$-maup
takes values close to zero when the variable is not sensitive to the MAUP and
values close to one when the variable is highly sensitive to the MAUP. Furthermore,
$S$-maup will be a univariate statistic applicable to spatially expansive
variables whose aggregated values result from the average of the individual
values.

\subsection{$S$-maup}\label{subsec:formu}

\begin{figure}
    \centering
	\includegraphics[width=0.99\textwidth]{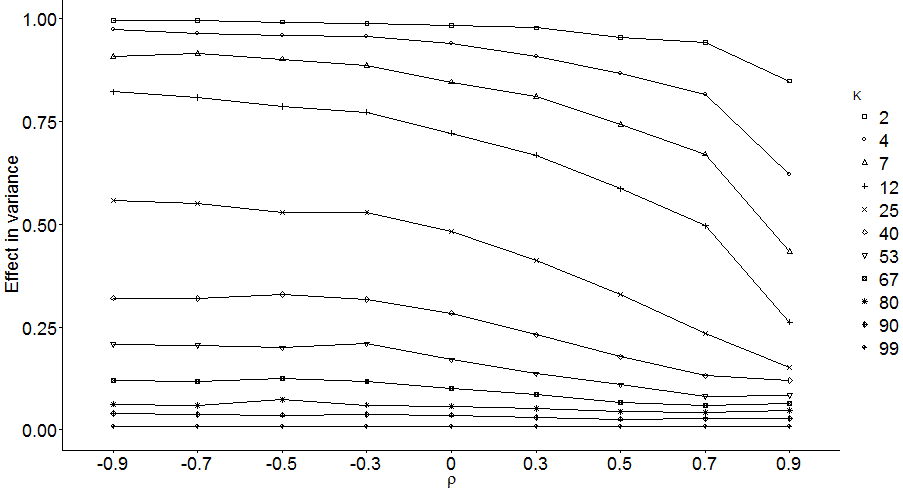}
	\caption{Median $\overline{RCM}$ for $N=100$.}
    \label{fig:efvl}
\end{figure}

To find the functional form of $S$-maup, it is necessary design an expression
that describes the distribution of the effects of MAUP on the variance
($\overline{RCV}$). To summarize those effects, we took the median of each Box
Plot in Fig \ref{fig:efv}. Fig \ref{fig:efvl} shows an example of those
summarized effects.

According to Fig \ref{fig:efvl}, the mathematical expression of our test
should take values close to one when the variable under evaluation has high
negative spatial autocorrelation, $\rho$ and is aggregated into a small
number of regions, $k$. Conversely, the expression should take values close to
zero when the variable under evaluation has high positive spatial
autocorrelation, $\rho$ and is aggregated into a large number of regions,
$k$. Our expression should also be able to reproduce the way in which, for a
give $k$, the MAUP effects decreases as $\rho$ increases. Note that such
a decrease is not the same for all values of $k$: when $k$ is large, the
effects of MAUP are low even for highly negative values of $\rho$; therefore,
for a high $k$, the reduction of the MAUP effects, as $\rho$ increases, are
almost imperceptible. Thus, as $k$ increases, our expression should modify
the speed and moment at which the MAUP fades along $\rho$. Taking into account
these different conditions, we started the construction of our $S$-maup
statistic using an inverted logistic function \cite{verhulst:1845}, which is
defined by Eq \eqref{eq:5}.

\begin{equation}\label{eq:5}
M(\rho; L, \eta, \tau  )= \frac{L}{1+\eta e^{\tau x}}   \  \ , \
\end{equation}
\\

\noindent where $L$ determines the maximum value of the curve; $\eta$
determines the moment at which the curve begins to decline; and $\tau$
indicates the speed at which the curve declines. If we endogenize those three
parameters, we should be able to approximate any line of the type shown in
Fig \ref{fig:efvl}. This is what we are going to develop in the rest of
this subsection until we obtain an expression of $M$ in which parameters
$L$, $\eta$ and $\tau$ depend on $\rho$, $k$ and $N$.

\begin{figure}
    \centering
	\includegraphics[width=0.99\textwidth]{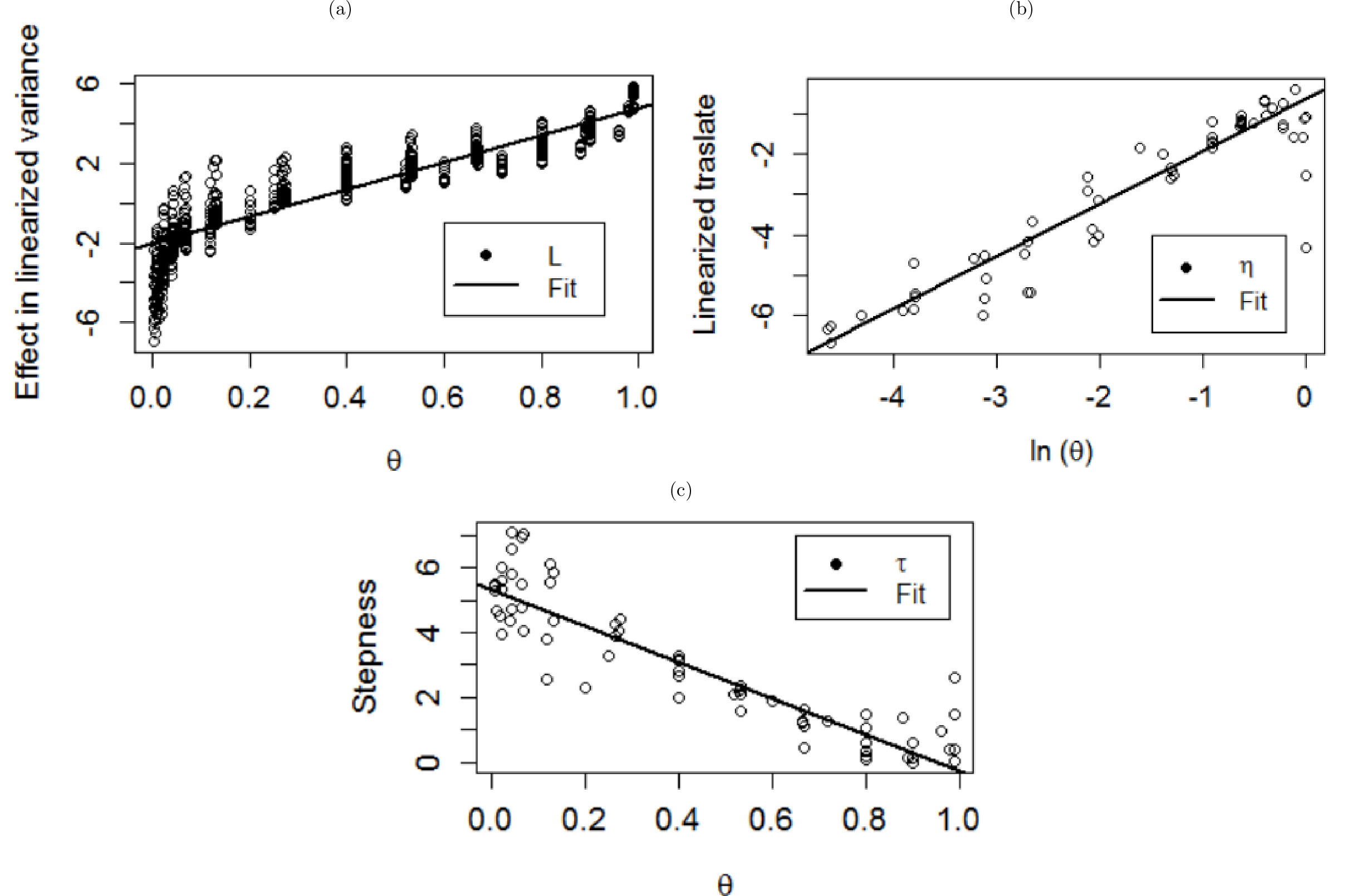}
	\caption{Adjustments of robust linear regression models: (a) Linearized
    logistic function ($L$); (b) Linearized power function ($\eta$); (c)
    Linear function ($\tau$).}
    \label{fig:calibra}
\end{figure}

Starting with the parameter $L$, Fig \ref{fig:efvl} shows that the maximum
value of each logistic curve depends on the level of aggregation $k$. This
aggregation can be defined in relative terms as $\theta=\frac{k}{N}$. Therefore, the lower
the level of aggregation (i.e., as $\theta$ approaches 1), the lower should be
$L$. When plotting each median $\overline{RVC}$ against $\theta$, it depicts an
inverted "S" that could also be modeled as an inverse logistic function with
the expression presented in Eq \eqref{eq:7}, whose linear form is given
by Eq \eqref{eq:7L}.

\begin{equation}\label{eq:7}
L( \theta )= \frac{1}{1+ e^{b+m\theta}}   \  \ , \
\end{equation}

\begin{equation}\label{eq:7L}
Ln \left (\frac{1-L}{L}\right )= b+m\theta  \  \ ,
\end{equation}

\noindent where $b$ and $m$ are the parameters of the inverse logistic
function. To estimate those parameters, we used a robust linear regression model that
minimizes the influence of outliers. The parameter associated with $\theta$ is
significant, and the adjusted R-squared = 86.7\%. Fig \ref{fig:calibra}(a)
shows the robust regression over the linearized logistic function.

Returning to the logistic curves in Fig \ref{fig:efvl}, both the moment at
which the curves begin to decrease, $\eta$, and the speed of decreasing,
$\tau$, depend on $k$. Therefore, both parameters can be estimated as function
of $\theta=\frac{k}{N}$. For this, we adjusted an inverse logistic function
for each curve of the type presented in Fig \ref{fig:efvl}. For each curve,
the values of $\eta$ and $\tau$ were calibrated using the optimized module of
Scipy Phyton Library \cite{jones:2001}. With this process, we obtained a
value for $\eta$ and $\tau$ for each value of $\theta$. Then, we use a
linearized power function, Eq \eqref{eq:8}, and a linear function,
Eq \eqref{eq:9}, to express $\eta$ and $\tau$ as a function of $\theta$.

\begin{equation}\label{eq:8}
\eta(\theta)= p\theta^{a} \ \
\end{equation}

\begin{equation}\label{eq:9}
\tau(\theta)= \beta_0 + \beta_1 \theta \ \
\end{equation}

The parameter associated with $\theta$ was significant in both estimations and
the adjusted $R^2$, with $91.7\%$ and $84.5\%$ respectively. Fig
\ref{fig:calibra}b and \ref{fig:calibra}c present the estimations.

Replacing the Eq \eqref{eq:7}, \eqref{eq:8} and \eqref{eq:9} in
\eqref{eq:5} we have the Eq \eqref{eq:10}.

\begin{equation}\label{eq:10}
M(\rho, \theta )= \frac{\frac{1}{1+ e^{b+m\theta}} }{1+p\theta^{a} e^{(\beta_0 + \beta_1 \theta) \rho}} \ \
\end{equation}

The results of the estimation of the parameters in the robust linear
regression model for the logistic function of $L$ are as follows: $m=7.031$ and
$b=-2.188$. Considering that the model is estimated with the linearized
logistic function, these results were transformed by natural logarithm. For
the power function of $\eta$, the results are $p=0.516$ and $a=1.287$,
because of the linearization of the power function, we applied the natural logarithm
to the parameter $p$. Finally, the results of the linear function of $\tau$ are as follows:
$\beta_0=5.319$ and $\beta_1=-5.532$. Replacing in the equations produces the following:

\begin{equation}\label{eq:11}
L( \theta )= \frac{1}{1+ e^{-2.188+7.301\theta}} \ \
\end{equation}

\begin{equation}\label{eq:12}
\eta(\theta)= 0.516\theta^{1.287} \ \
\end{equation}

\begin{equation}\label{eq:13}
\tau(\theta)= 5.319-5.532\theta.  \ \
\end{equation}
\\
Thus, the expression of the $S$-maup statistic is the following:\\
\ \
\begin{equation}\label{eq:14}
M(\rho, \theta)= \frac{\frac{1}{1+ e^{-2.188+7.031\theta}} }{1+ [0.516\theta^{1.287}] e^{[5.319-5.532\theta] \rho }} \ \
\end{equation}\\

Recall that $S$-maup statistic ($M$) is designed in such a way that for a
bigger (smaller) sensitivity of a variable to the MAUP, the larger (smaller) is
the value of $M$. This characteristic allows us to define a non-parametric
unilateral statistical test, which is stated below:

{\it $H_0$: The variable $y_i$ is not significantly affected by the MAUP.}\\

{\it $H_1$: The variable $y_i$ is significantly affected by the MAUP.}

\noindent Where the statistic for the test is given by Eq \eqref{eq:14},
and therefore, $H_0$ will be rejected if the statistic value belongs to the
rejection region ($RR$) defined in Eq \eqref{eq:16}.

\begin{equation}\label{eq:16}
RR=\left \{ M|M>M_{\alpha;\rho, N} \right \} \ \
\end{equation}

$M_{\alpha;\rho, N}$ is the critical value given a significance level
$\alpha$, a level of spatial autocorrelation ($\rho$), and a number of areas
($N$). We implemented a Monte Carlo simulation to find the empirical
distribution of the $S$-maup under the null hypothesis previously stated. The
empirical distribution allows us to obtain the critical values as well as the
pseudo-value $p$ to determine the proof significance.

\subsection{Critical values and $p$-value}\label{subsec:vc}

To calculate the critical values, we performed an exhaustive
simulation study based on non-parametric statistic methodology. Recall that
$H_0$ means no sensitivity of a variable to MAUP, which is equivalent to stating
that, for a given $k$, the variance of the aggregated variable is
statistically equal to the variance of the original variable. For building the
empirical distribution under $H_0$, we set a value for $N$ and $\rho$ and
generated an SAR process with parameters ($N$, $\rho$). Then, we randomly selected an integer value $k$ such that $0.1N < k < N$, thus yielding 30 random
aggregations of the variable into $k$ regions. Next, we applied the Levene test for equality of
variances between the original variable and each one of the 30 aggregated
variables. The SAR($N$, $\rho$) variable was kept if and only if the Levene test
was not rejected in all 30 cases. If there was at least one
rejection, then we chose, at random, a new $k$ and repeated the previous steps.
This procedure was repeated until we obtained 1,000 instances for each pair $(N,\rho)$.
We then calculated the $S$-maup statistic for those instances using Eq
\eqref{eq:14} and generated the empirical distribution of the
statistics under $H_0$. The critical values were obtained by calculating the
$90\%,\;\;95\%,\;\;99\%$ percentiles for the empirical distribution. Table
\ref{tab:vc} presents the table of critical values. This Table implied the
generation of 54,000 instances.

\begin{table}
    \centering
    \caption{{\bf Critical Values ($M_{\alpha;\rho, N}$)}}
    \begin{tabular}{cccccccc}
    \toprule
    & \multicolumn{7}{c}{Number of areas ($N$)}  \\
    $\rho$		&		$\alpha$		&    25  & 100  & 225  & 400 & 625 & 900 \\
    \midrule
    \multirow{3}{*}{-0.9} & 0.01 & 0.83702	& 0.09218 &	0.23808 &	0.05488 &	0.07218 &	0.02621 \\
    & 0.05 & 0.83699 &	0.08023 &	0.10962 &	0.04894 &	0.04641 &	0.02423 \\
    & 0.1  & 0.69331 &	0.06545 &	0.07858 &	0.04015 &	0.03374 &	0.02187 \\
    \midrule
    \multirow{3}{*}{-0.7} & 0.01 & 0.83676 &	0.16134 &	0.13402 &	0.06737 &	0.05486 &	0.02858 \\
    & 0.05 & 0.83662 &	0.12492 &	0.08643 &	0.05900 &	0.04280 &	0.02459 \\
    & 0.1  & 0.79421 &	0.09566 &	0.06777 &	0.05058 &	0.03392 &	0.02272 \\
    \midrule
    \multirow{3}{*}{-0.5} & 0.01 & 0.83597 &	0.16524 &	0.13446 &	0.06616 &	0.06247 &	0.02851 \\
    & 0.05 & 0.83578 &	0.13796 &	0.08679 &	0.05927 &	0.04260 &	0.02658 \\
    & 0.1  & 0.68900 &	0.10707 &	0.07039 &	0.05151 &	0.03609 &	0.02411 \\
    \midrule
    \multirow{3}{*}{-0.3} & 0.01 & 0.83316 &	0.19276 &	0.13396 &	0.06330 &	0.06090 &	0.03696 \\
    & 0.05 & 0.78849 &	0.16932 &	0.08775 &	0.05464 &	0.04787 &	0.03042 \\
    & 0.1  & 0.73592 &	0.14282 &	0.07076 &	0.04649 &	0.04001 &	0.02614 \\
    \midrule
    \multirow{3}{*}{0.0} & 0.01 & 0.82370 &	0.17925 &	0.15514 &	0.07732 &	0.07988 &	0.09301 \\
    & 0.05 & 0.81952 &	0.15746 &	0.11126 &	0.06961 &	0.06066 &	0.05234 \\
    & 0.1  & 0.71632 &	0.13621 &	0.08801 &	0.06112 &	0.04937 &	0.03759 \\
    \midrule
    \multirow{3}{*}{0.3} & 0.01 & 0.76472 &	0.23404 &	0.24640 &	0.11588 &	0.10715 &	0.07070 \\
    & 0.05 & 0.70466 &	0.21088 &	0.15360 &	0.09766 &	0.07938 &	0.06461 \\
    & 0.1  & 0.63718 &	0.18239 &	0.12101 &	0.08324 &	0.06347 &	0.05549 \\
    \midrule
    \multirow{3}{*}{0.5} & 0.01 & 0.67337 &	0.28921 &	0.25535 &	0.13992 &	0.12975 &	0.09856 \\
    & 0.05 & 0.59461 &	0.23497 &	0.18244 &	0.11682 &	0.10129 &	0.08860 \\
    & 0.1  & 0.46548 &	0.17541 &	0.14248 &	0.10008 &	0.08137 &	0.07701 \\
    \midrule
    \multirow{3}{*}{0.7} & 0.01 & 0.52155 &	0.47399 & 0.29351 &	0.23923 &	0.20321 &	0.16250 \\
    & 0.05 & 0.48958 &	0.37226 &	0.22280 &	0.20540 &	0.16144 &	0.14123 \\
    & 0.1  & 0.34720 &	0.28774 &	0.18170 &	0.16442 &	0.13395 &	0.12354 \\
    \midrule
    \multirow{3}{*}{0.9} & 0.01 & 0.28599 &	0.28938 &	0.43520 &	0.44060 &	0.34437 &	0.55967 \\
    & 0.05 & 0.21580 &	0.22532 &	0.27122 &	0.29043 &	0.23648 &	0.31424 \\
    & 0.1  & 0.17640 &	0.18835 &	0.21695 &	0.23031 &	0.19435 &	0.22411 \\
    \bottomrule
    \end{tabular}
    \label{tab:vc}
\end{table}

Following the percentile approach utilized by \cite{rey:2004}, we can
calculate a pseudo-$p$-value for a given value of the $S$-maup test ($M$),
using the Eq \eqref{eq:18}:

\begin{equation}\label{eq:18}
P(M)=  \frac{1}{1,000} \sum_{j=1}^{1,000} \Psi \  \ , \
\end{equation}

\noindent where $\Psi=1$ if $M_{j}^{\rho, N}>M$, $\Psi=0$ otherwise. The
vector $M_{j}^{\rho, N}$ comes from the simulations performed to produce Table
\ref{tab:vc}. Since those vectors are extremely computationally intensive to
produce (in some instances requiring months of supercomputer computation for
completion), they will be publicly available at http://www.\_\_\_.edu, as well
as the Python script to run the $S$-maup statistic.

Table \ref{tab:Smaup} presents some examples of the $S$-maup statistic for
different values of $N$ and $k$. Note that when the variable $y_i$ presents
characteristics against the null hypothesis ($H_0$), then the $M$ value of the
$S$-maup should be greater than the critical value at some significance level
$\alpha$, and therefore, the pseudo-value $p$ of the test must be smaller
than the significance level. If $H_0$ is rejected, it can be concluded that
the variable $y_i$ is sensitive to the MAUP, and therefore, a MAUP
effect exists when aggregating $y_i$ in $k$ regions.

\begin{table}
    \centering
    \caption{{\bf Example $S$-maup. }}
    \begin{tabular}{ccccccl}
    \toprule
    $Variable$		&		$N$		&  $k$   & $\rho$  & $M$  & $M_{\alpha;\rho, n}$ &Pseudo-v $p$\\
    \midrule
    $y_{i}^{1}$  & 1,000 & 400 & 0.007  & 0.24002 & 0.05234 & 0.0 ***  \\
    $y_{i}^{2}$  & 1,000 & 600 & 0.007  & 0.05871 & 0.05234 & 0.034 **   \\
    $y_{i}^{3}$  & 1,000 & 800 & 0.007  & 0.01187 & 0.05234 & 0.616     \\
    \midrule
    $y_{i}^{4}$  & 500  & 100 & -0.634 & 0.09237 & 0.05900 & 0.0 ***  \\
    $y_{i}^{5}$  & 500  & 280 & -0.634 & 0.05466 & 0.05900 & 0.078 *   \\
    $y_{i}^{6}$  & 500  & 380 & -0.634 & 0.00767 & 0.05900 & 0.852      \\
    \midrule
    $y_{i}^{7}$ & 220  & 60  & 0.562  & 0.32197 & 0.18244 & 0.00 *** \\
    $y_{i}^{8}$  & 220  & 90 & 0.562  & 0.18513 & 0.18244 & 0.046 **     \\
    $y_{i}^{9}$  & 220  & 150 & 0.562  & 0.04357 & 0.18244 & 0.443     \\
    \midrule
    $y_{i}^{10}$ & 150  & 15  & 0.801  & 0.29201  & 0.22532 & 0.009 **     \\
    $y_{i}^{11}$ & 150  & 50  & 0.801  & 0.08072 & 0.22532 & 0.366     \\
    $y_{i}^{12}$ & 150  & 90 & 0.801  & 0.00997 & 0.22532 & 0.883  	\\
    \bottomrule
    \end{tabular}
    \begin{flushleft}\footnotesize{*** $p < 0.01$, ** $p < 0.05$, * $p < 0.1$.}
    \end{flushleft}
    \label{tab:Smaup}
\end{table}

Note that when the spatial autocorrelation is highly positive
(e.g., $\rho=0.801$), the variable allows high levels of aggregation. The
results also confirm that low levels of spatial aggregation do not lead to
the undesirable effects of MAUP.

\section{Power and Size}
\label{sec:prop}

The power is a natural way of evaluating the test performance. It is defined
as the probability of rejecting the null hypothesis, given that the null
hypothesis is false. In other words, it is the probability of not committing
a type II error ($\beta$); thus, the power is ($1-\beta$). In our context, the
power means the probability that sufficient statistical evidence exists in the
sample to affirm that the variable $y_i$ is affected by the MAUP, when in
fact, the variable $y_i$ is affected by the dimensions of the MAUP. Hence, it
is expected that the power of the test is close, or equal, to 1.

Since $H_1$ implies that the variance of the original variable is different
from the variance of the aggregate variable, we implemented the following
simulation experiment to measure the power of our statistical test: For each
tuple $(N,\rho)$, with $N \in \left \{100,400,900 \right \}$ and $\rho \in
\left \{ \pm 0.9, \pm 0.7, \pm 0.5, \pm 0.3, 0 \right \}$. Given a
tuple $(N,\rho)$ we generate an SAR process and perform 30 random spatial
aggregations of the $N$ areas into $k$ regions such that $k$ is selected at
random as an integer value such that $0.1N < k < N$. The SAR process is kept if
and only if the Leven test between the original variable and each one of the
30 aggregated variables is rejected. We repeat this process until we generate
1,000 valid instances for each tuple $(N,\rho)$. Each entry in Table
\ref{tab:pot} reports the proportion of 1,000 instances for which our test
rejects $H_0$. Because most values are close to 1, we can argue that our $S$-maup is
highly effective in identifying variables that are sensitive to
the MAUP effect.

\begin{table}
    \centering
    \caption{{\bf Estimated power of $S$-maup.}}
    \begin{tabular}{cccc}
    \toprule
    & \multicolumn{3}{c}{Number of areas ($N$)}  \\
    $\rho$& $N=100$ & $N=400$ & $N=900$ \\
    \midrule
    -0.9 & 0.989 & 0.985 & 0.997 \\
    -0.7 & 0.986 & 0.996 & 1.000 \\
    -0.5 & 0.981 & 0.998 & 1.000 \\
    -0.3 & 0.982 & 0.998 & 1.000 \\
    0.0  & 0.997 & 0.999 & 0.999 \\
    0.3  & 0.986 & 0.996 & 1.000 \\
    0.5  & 0.986 & 0.996 & 0.999 \\
    0.7  & 0.783 & 0.985 & 0.995 \\
    0.9  & 0.977 & 0.703 & 0.492 \\
    \bottomrule
    \end{tabular}
    \begin{flushleft}\footnotesize{Level of significance $\alpha=0.05$.}
    \end{flushleft}
    \label{tab:pot}
\end{table}

Test size is also a way of evaluating the test performance. Test size is defined as
the probability of rejecting the null hypothesis given that the null
hypothesis is true. In other words, it is the probability of committing a
type I error ($\alpha$). In our context, test size means the probability that
sufficient statistical evidence exists in the sample to affirm that the
variable $y_i$ is affected by the MAUP, when in fact the variable $y_i$ is
not. Hence, it is expected that the proportion of instances for which our test
commits type I error is close the theoretical significance level ($\alpha$).

The empirical test size is calculated following a similar procedure
implemented to calculate the power, but in this case, the tuple $(N, \rho)$ is
selected if and only if the Levene test is not rejected in all 30 cases. Table
\ref{tab:ta} reports the size of our test, which show the best performance in
scenarios of positive spatial autocorrelation.

\begin{table}
    \centering
    \caption{{\bf Estimated size of $S$-maup.}}
    \begin{tabular}{cccc}
    \toprule
    & \multicolumn{3}{c}{Number of areas ($N$)}  \\
    $\rho$& $N=100$ & $N=400$ & $N=900$ \\
    \midrule
    -0.9 & 0.163 & 0.087 & 0.065 \\
    -0.7 & 0.080 & 0.037 & 0.080 \\
    -0.5 & 0.091 & 0.043 & 0.083 \\
    -0.3 & 0.073 & 0.097 & 0.136 \\
    0    & 0.102 & 0.066 & 0.026 \\
    0.3  & 0.081 & 0.057 & 0.038 \\
    0.5  & 0.098 & 0.062 & 0.032 \\
    0.7  & 0.043 & 0.032 & 0.045 \\
    0.9  & 0.110 & 0.024 & 0.009 \\
    \bottomrule
    \end{tabular}
    \begin{flushleft}\footnotesize{Level of significance $\alpha=0.05$.}
    \end{flushleft}
    \label{tab:ta}
\end{table}

\section{An illustrative application of the $S$-maup test}
\label{sec:caso}

In this section, we present an empirical illustration within the context of a
Mincer wage equation \cite{Mincer:1974} that explains the salary based on
schooling and experience. Eq \eqref{eq:mincer} presents the most basic
version of the Mincer wage equation.

\begin{equation}\label{eq:mincer}
LNW=  \beta_0 + \beta_1 * YRSCHOOL + \beta_2 * EXP + \beta_3 * EXP^2 + \varepsilon , \
\end{equation}

\noindent where $LNW$ is the natural logarithm of income (hourly wage),
$YRSCHOOL$ years of schooling, $EXP$ years of potential labor market
experience (calculated as the age in years minus years of education plus 6),
and $\varepsilon$ is a mean zero residual. It is important to clarify that
this example is merely illustrative.  We use this equation because its
simplicity allows us to present a simple application of our test.

We use the 2011 census data from South Africa retrieved from the Integrated
Public Use Microdata Series, International (IPUMS-International), at the
Minnesota Population Center \cite{center:2015}. The data include 688,310
individuals who were working at the time of the survey. We
aggregate the individual data into 206 municipalities using the weighted average
of individual incomes, years of schooling, and the potential work experience.

\begin{figure}
    \centering
	\includegraphics[width=0.99\textwidth]{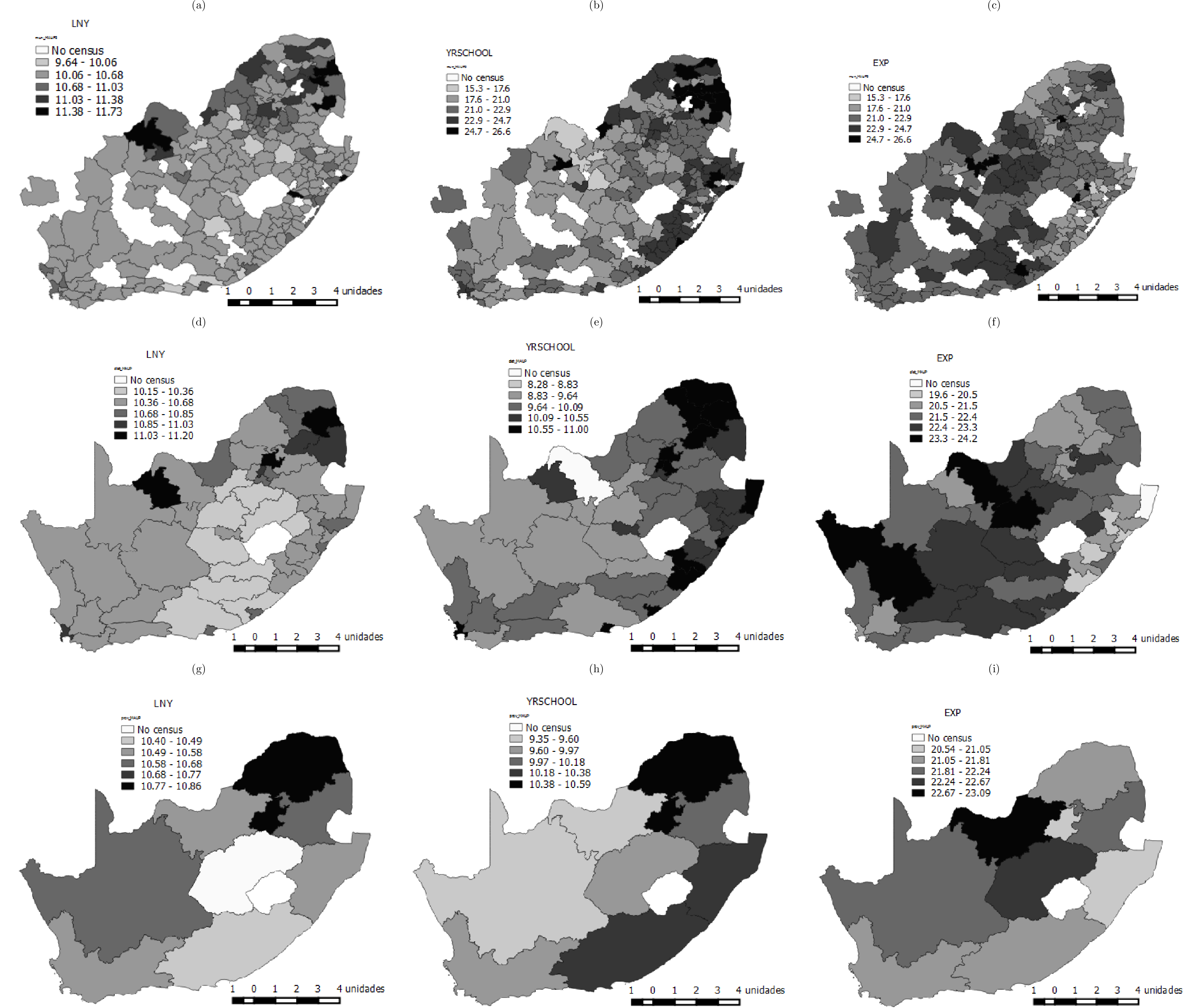}
	\caption{Municipalities: (a), (b) and (c). Districts: (d), (e) and (f).
    Provinces:(g), (h) and (i).}
    \label{fig:mun}
\end{figure}

The 206 municipalities are our basic unit of analysis (i.e., our disaggregated
variable). Other administrative units in South Africa include 52 districts and
9 provinces. Table \ref{tab:descip} shows some descriptive statistics of our
variables at the three administrative levels. Note how the standard deviation
of the three variables narrows as the level of aggregation increases. The
spatial distribution of the variables is presented in Fig
\ref{fig:mun}.

\begin{table}
    \centering
    \caption{{\bf Descriptive Statistics.}}
    \begin{tabular}{lccccc}
    \toprule
    & \multicolumn{5}{c}{Municipalities}  \\
    Variable	&	Obs.& Mean  & Desv. Std.& Mín.  & Máx. \\
    \midrule
    LNW  			& 206 & 10.51 & 0.35  & 9.64 & 11.73   \\
    YRSCHOOL  & 206 & 9.95 & 0.81  & 7.43 & 11.87    \\
    EXP  			& 206 & 21.69 & 1.71  & 15.28 & 26.64  \\
    \midrule
    & \multicolumn{5}{c}{Districts}  \\
    \midrule
    LNW  			& 52 & 10.56 & 0.25  & 10.15 & 11.20   \\
    YRSCHOOL  & 52 & 10.06 & 0.61  & 8.28 & 10.99    \\
    EXP  			& 52 & 21.59 & 1.21  & 18.66 & 24.24  \\
    \midrule
    & \multicolumn{5}{c}{Provinces}  \\
    \midrule
    LNW  			& 9 & 10.57 & 0.19  & 10.31 & 10.86   \\
    YRSCHOOL  & 9 & 10.00 & 0.44  & 9.35 & 10.59    \\
    EXP  			& 9 & 21.77 & 0.77  & 20.54 & 23.09  \\
    \bottomrule
    \end{tabular}
    \label{tab:descip}
\end{table}

Table \ref{tab:mod} presents the estimation at the municipal level. The
coefficients of education and experience are significant and exhibit the
expected signs.

\begin{table}
    \centering
    \caption{{\bf Mincer Model Estimate: South Africa.}}
    \begin{tabular}{lcclcc}
    \toprule
    LNW				&	Coef.& Desv. Std  & $p>|t|$  & \multicolumn{2}{c}{Confidence Interval at 95\%}  \\
    \midrule
    YRSCHOOL  & 0.3364 & 0.0259 & 0.000 *** & 0.2852 & 0.3876    \\
    EXP  			& 0.4008 & 0.1499 & 0.008 *** & 0.1051 & 0.6965  \\
    EXP2  			& -0.0085 & 0.0034 & 0.016 ** & -0.0153 & -0.0016  \\
    CONST.			& 2.4796 & 1.6243 & 0.128  & -0.7232 & 5.6825  \\
    \midrule
    \multicolumn{6}{l}{Num. Obs. 206}  \\
    \multicolumn{6}{l}{F(3,202) = 68.84}  \\
    \multicolumn{6}{l}{$R^{2}$ adjusted = 0.498}  \\
    \bottomrule
    \end{tabular}
    \begin{flushleft}\footnotesize{*** $p < 0.01$, ** $p < 0.05$, * $p < 0.1$.}
    \end{flushleft}
    \label{tab:mod}
\end{table}

What would be the maximum level of spatial aggregation for which these results
hold? Note that here we are asking about the minimum value for $k$ that
preserves the distributional characteristics of the variables; we are not
aiming to evaluate a specific regionalization for a give value of $k$. We can
use our $S$-maup statistic to answer this question by identifying the minimum
value of $k$ for which our test fails to reject the null hypothesis of no
influence of the MAUP. In Table \ref{tab:resulSmaup}, we present the results of
our test for different levels of spatial aggregations. For this, our test
requires the level of spatial autocorrelation of each variable ($\rho$) and
the value of $\theta=\frac{k}{N}$. Note that at $k=135$, the $S$-maup indicates
that the variable $LNW$ is affected by the MAUP. This finding may imply that the
results obtained at municipal level ($k=206$) may hold until an aggregation
level of $k=136$ that is the aggregation level at which all the variables
involved in the regression do not lose their distributional characteristics.
Another conclusion from these results is that the results obtained at the
municipal level do not hold at district or province levels.

\begin{table}
    \centering
    \caption{{\bf Estimator of the statistic $ S $ -maup: South Africa.}}
    \begin{tabular}{lclclclcl}
    \toprule
        & \multicolumn{8}{c}{$N=206$} \\
        & \multicolumn{2}{c}{LNW} & \multicolumn{2}{c}{YRSCHOOL} & \multicolumn{2}{c}{EXP} & \multicolumn{2}{c}{EXP2} \\
        & \multicolumn{2}{c}{$\rho=0.05$} & \multicolumn{2}{c}{$\rho=0.25$} & \multicolumn{2}{c}{$\rho=0.24$} & \multicolumn{2}{c}{$\rho=0.40$} \\
    \midrule
    $k$  & $M$  & Ps-v $p$ & $M$  & Ps-v $p$ & $M$  & Ps-v $p$ & $M$  & Ps-v $p$ \\
    \midrule
    200	 & 0.011 & 0.806 & 0.011 & 0.820 & 0.011 & 0.819 & 0.011 & 0.833 \\
    180	 & 0.022 & 0.589 & 0.021 & 0.619 & 0.021 & 0.619 & 0.020 & 0.656 \\
    150	 & 0.057 & 0.242 & 0.052 & 0.330 & 0.053 & 0.327 & 0.048 & 0.414 \\
		136  & 0.087 & 0.101 & 0.079 & 0.197 & 0.079 & 0.194 & 0.072 & 0.302 \\
		135  & 0.091 & 0.094 * & 0.081 & 0.187 & 0.082 & 0.185 & 0.073 & 0.295 \\
		134  & 0.093 & 0.089 * & 0.083 & 0.181 & 0.084 & 0.179 & 0.076 & 0.290 \\
		132  & 0.099 & 0.077 * & 0.088 & 0.166 & 0.089 & 0.166 & 0.079 & 0.273 \\
    124	 & 0.124 & 0.036 ** & 0.111 & 0.115 & 0.112 & 0.114 & 0.099 & 0.208 \\
    122  & 0.131 & 0.032 ** & 0.117 & 0.107 & 0.118 & 0.104 & 0.104 & 0.186 \\
    120  & 0.139 & 0.025 ** & 0.123 & 0.094 * & 0.125 & 0.091 * &  0.110 & 0.167 \\
    118  & 0.147 & 0.019 ** & 0.131 & 0.081 * & 0.132 & 0.080 * & 0.142 & 0.101 \\
    110  & 0.182 & 0.003 ** & 0.161 & 0.043 ** & 0.163 & 0.042 ** & 0.149 & 0.093 * \\
    108  & 0.192 & 0.001 ** & 0.169 & 0.034 ** & 0.172 & 0.033 ** & 0.149 & 0.093 * \\
    52   & 0.584 & 0.000 *** & 0.527 & 0.001 *** & 0.533 & 0.001 *** & 0.461 & 0.00 ***\\
    9		 & 0.863 & 0.000 *** & 0.847 & 0.001 *** & 0.849 & 0.001 *** & 0.822 & 0.00 ***\\
    \bottomrule
    \end{tabular}
    \begin{flushleft}\footnotesize{*** $p < 0.01$, ** $p < 0.05$, * $p < 0.1$.}
    \end{flushleft}
    \label{tab:resulSmaup}
\end{table}

\begin{figure}
    \centering
	\includegraphics[width=0.99\textwidth]{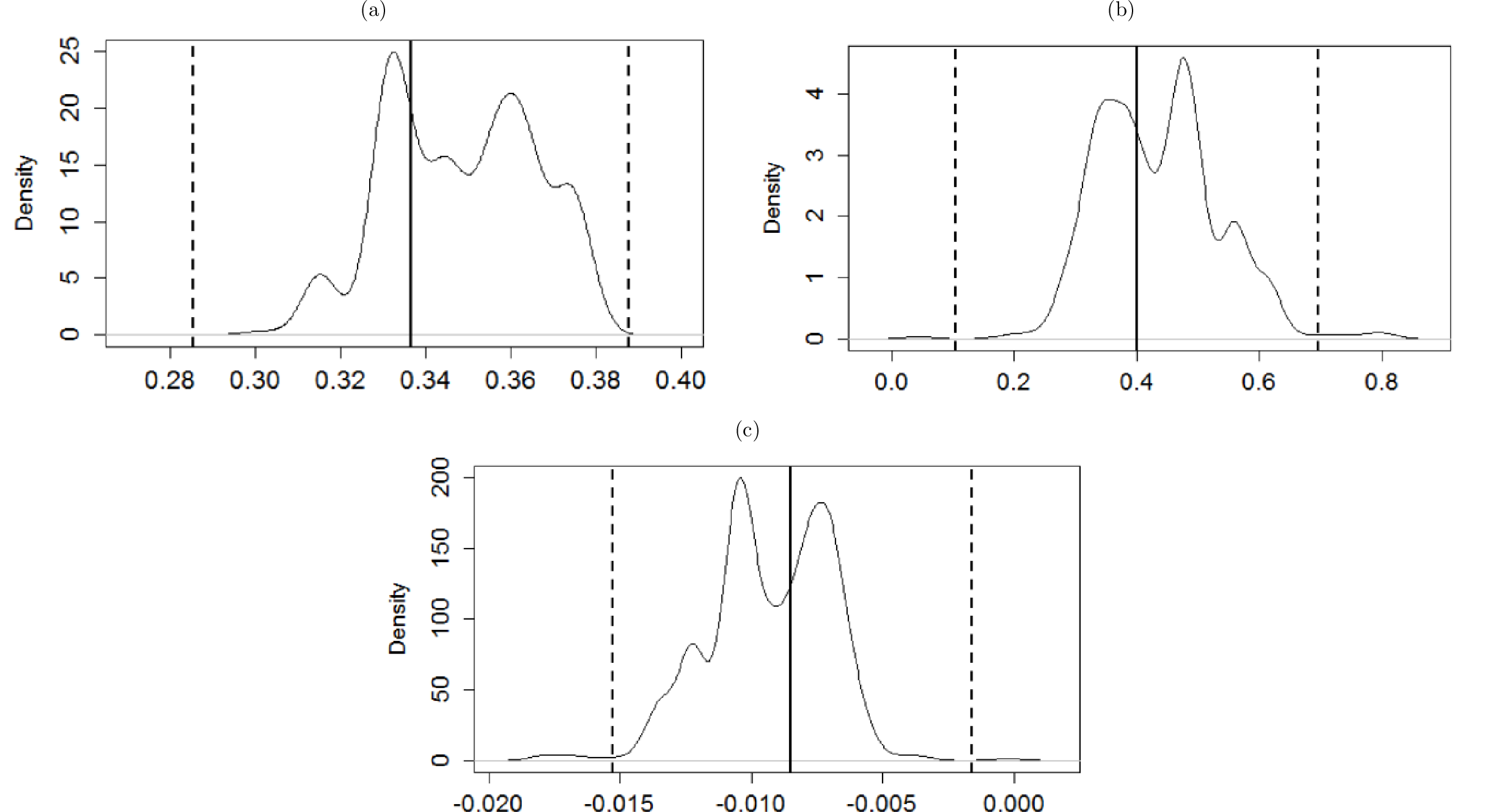}
    \caption{Distribution of coefficients, $k=136$: (a) YRSCHOOL; (b) EXP; (c)
    EXP2. horizontal black line: coefficient (206 municipalities), dashed
    lines are the respective confidence intervals 95\%.}
    \label{fig:ols}
\end{figure}

Fig \ref{fig:ols} compares the coefficients obtained at the municipal level
(black and dashed vertical lines) with the distribution of the coefficients
obtained by estimating the Mincer equation on 1,000 random spatial
aggregations of the $k=206$ municipalities into $k=136$ regions. Fig
\ref{fig:ols}a, corresponding to years of education, shows that 100\% of the
coefficients estimated with $k=136$ fall into the 95\% confidence intervals.
Fig \ref{fig:ols}b, corresponding to years of experience, shows that 98.8\%
of the coefficients estimated with $k=136$ fall into the 95\% confidence
intervals. Finally, Fig \ref{fig:ols}c, corresponding to the squared
years of experience, shows that 98.7\% of the coefficients estimated with
$k=136$ fall into the 95\% confidence intervals.

\begin{figure}
    \centering
	\includegraphics[width=0.99\textwidth]{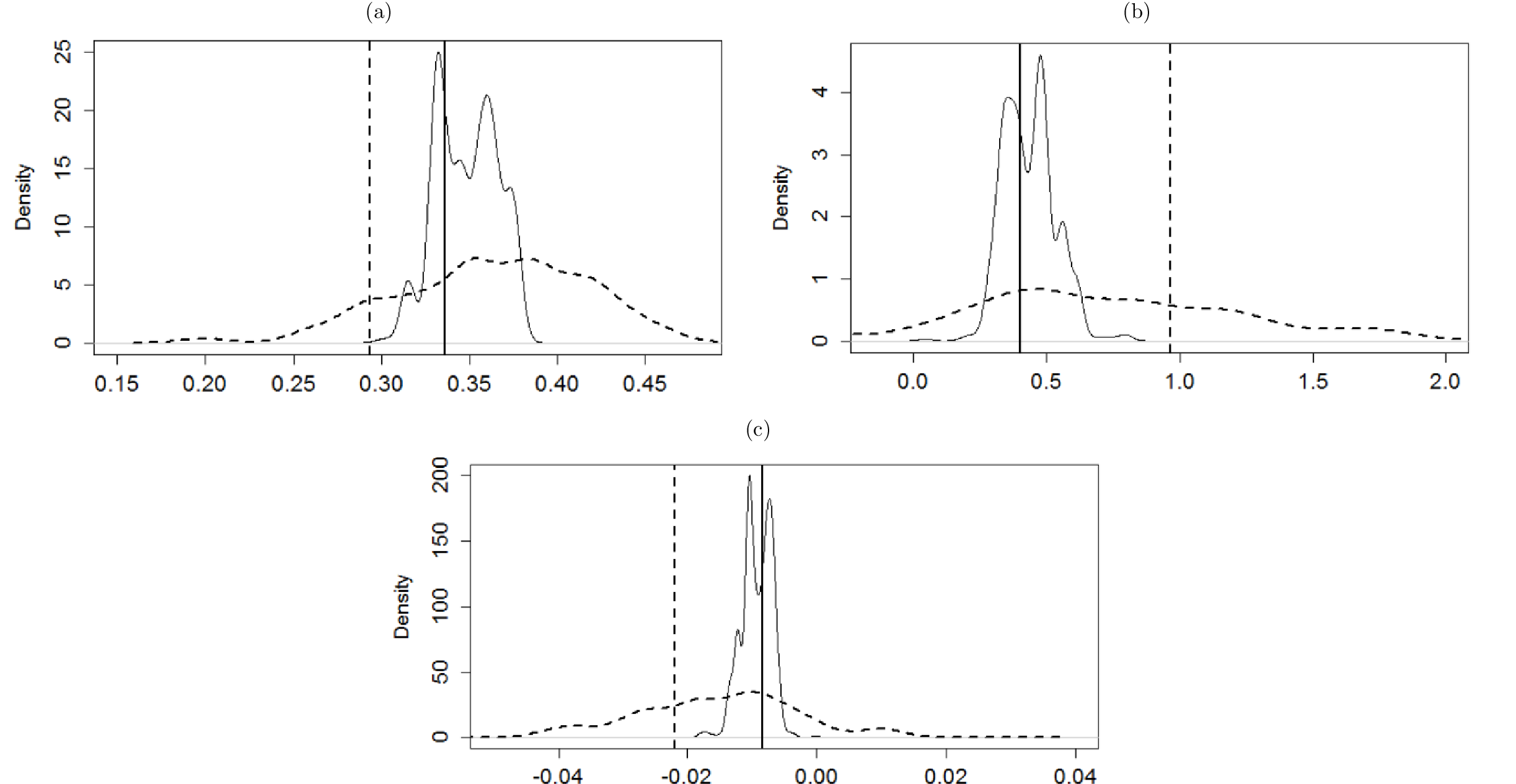}
    \caption{Distribution of coefficients. line:$k=136$, dotted line:$k=52$:
    (a) YRSCHOOL; (b) EXP; (c) EXP2. horizontal black line: coefficient (206
    municipalities). horizontal dotted line: coefficient (52 districts).}
    \label{fig:olsk52}
\end{figure}

Next, we estimated the Mincer model for $k=52$ and compared it with the
estimation for $k=206$ municipalities. As we did previously, we made 1,000
random aggregations and obtained the distribution of the estimated coefficients
for $K=136$ and $k=52$. Fig \ref{fig:olsk52} shows how the estimations with
$k=52$ are more volatile and deviated than those with $k=136$ regions. Note
also that the coefficients for $K=206$

\section{Conclusions}
\label{sec:conclu}

This paper introduced the first statistic of its kind for measuring the level
of sensitivity of a spatially expansive variable to the MAUP. The statistic
is easy to implement because it only requires as input parameters the level of
aggregation $\theta=\frac{k}{N}$ and the level of spatial autocorrelation of
the variable $\rho$. The test exhibits good statistical power and size. We also
provide the table of critical values and a procedure to calculate the
pseudo-$p$ value of the test.

The empirical application shows the usefulness of the test for identifying the
maximum level of aggregation at which the original variable preserves its
distributional characteristics. Additionally, it can be useful to test whether two
aggregation levels are comparable.

We recognize that the main properties of the $S$-maup were obtained from an
empirical simulation procedure, and they rely more heavily on hard
experimental computation than theoretical methods. However, the complexity of
the question addressed in this paper may explain why this is the first attempt
to answer it even though the MAUP has been in the literature since the late 1970s.
We hope that this first attempt motivates other researchers to contribute other
approaches to answer the same question.

\section*{Acknowledgement}\label{sec:acknowledgement}

We thank Professor Andrés Ramírez Hassan for his comments. We also thank the
Cyberinfrastructure Service for High Performance Computing, Apolo, at
Universidad EAFIT for letting us run our computational experiments on their
supercomputer.

{}


\begin{thebibliography}{}

\bibitem[Amrhein, 1995]{amrhein:1995}
Amrhein, C.~G. (1995).
\newblock Searching for the elusive aggregation effect: evidence from
  statistical simulations.
\newblock {\em Environment and planning A}, 27(1):105--119.

\bibitem[Amrhein and Reynolds, 1996]{amrheinRey:1996}
Amrhein, C.~G. and Reynolds, H. (1996).
\newblock Using spatial statistics to assess aggregation effects.
\newblock {\em Geographical Systems}, 3(2/3):143--158.

\bibitem[Arbia, 1989]{arbia:1989}
Arbia, G. (1989).
\newblock {\em Spatial data configuration in statistical analysis of regional
  economic and related problems}.
\newblock Dordrecht and kluwer academic, Boston.

\bibitem[Arbia et~al., 1996]{arbia:1996}
Arbia, G., Espa, G., et~al. (1996).
\newblock Effects of the maup on image classification.
\newblock {\em Geographical Systems}, (3):123--141.

\bibitem[Arbia and Petrarca, 2011]{arbia:2011}
Arbia, G. and Petrarca, F. (2011).
\newblock Effects of maup on spatial econometric models.
\newblock {\em Letters in Spatial and Resource Sciences}, 4(3):173--185.

\bibitem[Arbia and Petrarca, 2013]{arbia:2013}
Arbia, G. and Petrarca, F. (2013).
\newblock Effects of scale in spatial interaction models.
\newblock {\em Journal of Geographical Systems}, 15(3):249--264.

\bibitem[Bian and Butler, 1999]{bian:99}
Bian, L. and Butler, R. (1999).
\newblock Comparing effects of aggregation methods on statistical and spatial
  properties of simulated spatial data.
\newblock {\em Photogrammetric Engineering and Remote Sensing}, 65:73--84.

\bibitem[Carrington et~al., 2006]{carrington:06}
Carrington, A., Rahman, N., and Ralphs, M. (2006).
\newblock 11th meeting of the national statistics methodology advisory
  committee.

\bibitem[Center, 2015]{center:2015}
Center, M.~P. (2015).
\newblock Integrated public use microdata series, international: Version 6.4
  [database].
\newblock {\em University of Minnesota, Minneapolis.
  http://doi.org/10.18128/D020.V6.4.}

\bibitem[Clark and Avery, 1976]{clark:1976}
Clark, W.~A. and Avery, K.~L. (1976).
\newblock The effects of data aggregation in statistical analysis.
\newblock {\em Geographical Analysis}, 8(4):428--438.

\bibitem[Coulson, 1978]{coulson:1978}
Coulson, M.~R. (1978).
\newblock "potential for variation": A concept for measuring the significance
  of variations in size and shape of areal units.
\newblock {\em Geografiska Annaler. Series B. Human Geography}, pages 48--64.

\bibitem[Cressie, 1996]{cressie:1996}
Cressie, N.~A. (1996).
\newblock Change of support and the modifiable areal unit problem.

\bibitem[Duque et~al., 2006]{Duque2006}
Duque, J.~C., Art{\'{i}}s, M., and Ramos, R. (2006).
\newblock {The ecological fallacy in a time series context: Evidence from
  Spanish regional unemployment rates}.
\newblock {\em Journal of Geographical Systems}, 8(4):391--410.

\bibitem[Duque et~al., 2011]{ClusterPy}
Duque, J.~C., Dev, B., Betancourt, A., and Franco, J.~L. (2011).
\newblock {\em ClusterPy: {Library} of spatially constrained clustering
  algorithms, {Version} 0.9.9.}
\newblock RiSE-group (Research in Spatial Economics). EAFIT University.,
  Colombia.

\bibitem[Duque et~al., 2012]{Duque2012}
Duque, J.~C., Royuela, V., and Nore{\~{n}}a, M. (2012).
\newblock {A stepwise procedure to determinate a suitable scale for the spatial
  delimitation of urban slums}.
\newblock In {\em Advances in Spatial Science}, volume~75, pages 237--254.

\bibitem[Flowerdew and Amrhein, 1989]{flowerdewP:1989}
Flowerdew, R. and Amrhein, C. (1989).
\newblock Poisson regression models of canadian census division migration
  flows.
\newblock {\em Papers in Regional Science}, 67(1):89--102.

\bibitem[Fotheringham, 1989]{fotheringham:1989}
Fotheringham, A.~S. (1989).
\newblock {\em Scale-independent spatial analysis}, pages 221--228.
\newblock Taylor and Francis London, USA.

\bibitem[Fotheringham et~al., 2000]{fotheringham:2000}
Fotheringham, A.~S., Brunsdon, C., and Charlton, M. (2000).
\newblock {\em Quantitative geography: perspectives on spatial data analysis}.
\newblock Sage.

\bibitem[Fotheringham and Wong, 1991]{fotheringham:1991}
Fotheringham, A.~S. and Wong, D.~W. (1991).
\newblock The modifiable areal unit problem in multivariate statistical
  analysis.
\newblock {\em Environment and planning A}, 23(7):1025--1044.

\bibitem[Gehlke and Biehl, 1934]{gehlke:1934}
Gehlke, C.~E. and Biehl, K. (1934).
\newblock Certain effects of grouping upon the size of the correlation
  coefficient in census tract material.
\newblock {\em Journal of the American Statistical Association},
  29(185A):169--170.

\bibitem[Goodchild, 1979]{goodchild:1979}
Goodchild, M.~F. (1979).
\newblock The aggregation problem in location-allocation.
\newblock {\em Geographical Analysis}, 11(3):240--255.

\bibitem[Green and Flowerdew, 1996]{green:1996}
Green, M. and Flowerdew, R. (1996).
\newblock New evidence on the modifiable areal unit problem.
\newblock {\em Spatial analysis: Modelling in a GIS environment}, pages 41--54.

\bibitem[Guo and Bhat, 2004]{guo:2004}
Guo, J. and Bhat, C. (2004).
\newblock Modifiable areal units: Problem or perception in modeling of
  residential location choice?
\newblock {\em Transportation Research Record: Journal of the Transportation
  Research Board}, (1898):138--147.

\bibitem[Holt et~al., 1996a]{holtarea:1996}
Holt, D., Steel, D., and Tranmer, M. (1996a).
\newblock Area homogeneity and the modifiable areal unit problem.
\newblock {\em Geographical Systems}, 3(2/3):181--200.

\bibitem[Holt et~al., 1996b]{holt:1996}
Holt, D., Steel, D.~G., Tranmer, M., and Wrigley, N. (1996b).
\newblock Aggregation and ecological effects in geographically based data.
\newblock {\em Geographical Analysis}, 28(3):244--261.

\bibitem[Hunt and Boots, 1996]{hunt:1996}
Hunt, L. and Boots, B. (1996).
\newblock Maup effects in the principal axis factoring technique.
\newblock {\em Geographical Systems}, 3(2/3):101--122.

\bibitem[Jelinski and Wu, 1996]{jelinski:1996}
Jelinski, D.~E. and Wu, J. (1996).
\newblock The modifiable areal unit problem and implications for landscape
  ecology.
\newblock {\em Landscape ecology}, 11(3):129--140.

\bibitem[Jones et~al., 2001]{jones:2001}
Jones, E., Oliphant, T., Peterson, P., et~al. (2001).
\newblock Scipy: Open source scientific tools for python, 2009.
\newblock {\em URL http://scipy. org}.

\bibitem[King, 1997]{king:1997}
King, G. (1997).
\newblock A solution to the ecological inference problem.

\bibitem[Miller, 1999]{miller:99}
Miller, H.~J. (1999).
\newblock Potential contributions of spatial analysis to geographic information
  systems for transportation (gis-t).
\newblock {\em Geographical Analysis}, 31(4):373--399.

\bibitem[Miller, 1998]{miller:1998}
Miller, J.~R. (1998).
\newblock Spatial aggregation and regional economic forecasting.
\newblock {\em The Annals of Regional Science}, 32(2):253--266.

\bibitem[Mincer, 1974]{Mincer:1974}
Mincer, J. (1974).
\newblock Schooling, experience and earnings.
\newblock {\em National Bureau of Economic Research}.

\bibitem[Moellering and Tobler, 1972]{moellering:1972}
Moellering, H. and Tobler, W. (1972).
\newblock Geographical variances.
\newblock {\em Geographical Analysis}, 4(1):34--50.

\bibitem[Nakaya, 2000]{nakaya:2000}
Nakaya, T. (2000).
\newblock An information statistical approach to the modifiable areal unit
  problem in incidence rate maps.
\newblock {\em Environment and Planning A}, 32(1):91--109.

\bibitem[Openshaw, 1977]{openshaw:1977}
Openshaw, S. (1977).
\newblock A geographical solution to scale and aggregation problems in
  region-building, partitioning and spatial modelling.
\newblock {\em Transactions of the institute of british geographers}, pages
  459--472.

\bibitem[Openshaw, 1978]{openshaw:1978}
Openshaw, S. (1978).
\newblock An empirical study of some zone-design criteria.
\newblock {\em Environment and planning A}, 10(7):781--794.

\bibitem[Openshaw and Taylor, 1979]{openshaw:1979}
Openshaw, S. and Taylor, P.~J. (1979).
\newblock A million or so correlation coefficients: three experiments on the
  modifiable areal unit problem.
\newblock {\em Statistical applications in the spatial sciences}, 21:127--144.

\bibitem[Qi and Wu, 1996]{qi:1996}
Qi, Y. and Wu, J. (1996).
\newblock Effects of changing spatial resolution on the results of landscape
  pattern analysis using spatial autocorrelation indices.
\newblock {\em Landscape ecology}, 11(1):39--49.

\bibitem[Rey, 2004]{rey:2004}
Rey, S.~J. (2004).
\newblock Spatial analysis of regional income inequality.
\newblock {\em Spatially Integrated Social Science}, 1:280--299.

\bibitem[Reynolds, 1998]{reynolds:1998}
Reynolds, H.~D. (1998).
\newblock {\em The modifiable area unit problem: empirical analysis by
  statistical simulation}.
\newblock PhD thesis, Citeseer.

\bibitem[Robinson, 1950]{robinson:1950}
Robinson, W.~S. (1950).
\newblock Ecological correlations and the behavior of individuals.
\newblock {\em American Sociological Review}, 15(3):351--357.

\bibitem[Steel and Holt, 1996]{steelrules:1996}
Steel, D. and Holt, D. (1996).
\newblock Rules for random aggregation.
\newblock {\em Environment and Planning A}, 28(6):957--978.

\bibitem[Tagashira and Okabe, 2002]{tagashira:02}
Tagashira, N. and Okabe, A. (2002).
\newblock The modifiable areal unit problem, in a regression model whose
  independent variable is a distance from a predetermined point.
\newblock {\em Geographical analysis}, 34(1):1--20.

\bibitem[Verhulst, 1845]{verhulst:1845}
Verhulst, P.~F. (1845).
\newblock Recherches math{\'e}matiques sur la loi d'accroissement de la
  population.
\newblock {\em Nouveaux M{\'e}moires de l'Acad{\'e}mie Royale des Sciences et
  Belles-Lettres de Bruxelles}, 18:14--54.

\bibitem[Vickrey, 1961]{vickrey:1961}
Vickrey, W. (1961).
\newblock On the prevention of gerrymandering.
\newblock {\em Political Science Quarterly}, 76(1):105--110.

\bibitem[Wise et~al., 1997]{wise:97}
Wise, S., Haining, R., and Ma, J. (1997).
\newblock {\em Regionalisation tools for the exploratory spatial analysis of
  health data}.
\newblock Springer.

\bibitem[Wise et~al., 2001]{wise:01}
Wise, S., Haining, R., and Ma, J. (2001).
\newblock Providing spatial statistical data analysis functionality for the gis
  user: The sage project.
\newblock {\em International Journal of Geographical Information Science},
  15(3):239--254.

\bibitem[Wrigley et~al., 1996]{wrigleyb:1996}
Wrigley, N., Holt, T., Steel, D., and Tranmer, M. (1996).
\newblock {\em Analysing, Modelling, and Resolving the Ecological Fallacy},
  pages 23--40.
\newblock John Wiley and Sons, New York.

\bibitem[Yule and Kendall, 1950]{yule:1950}
Yule, G.~U. and Kendall, M. (1950).
\newblock An introduction to the theory of statistics.
\newblock {\em some measures of status inconsistency}.

\end{thebibliography}

\end{document}